\documentclass{article}

\pdfoutput=1
\usepackage{PRIMEarxiv}

\usepackage[utf8]{inputenc} 
\usepackage[T1]{fontenc}    
\usepackage{hyperref}       
\usepackage{url}            
\usepackage{booktabs}       
\usepackage{amsfonts}       
\usepackage{nicefrac}       
\usepackage{microtype}      
\usepackage{lipsum}
\usepackage{fancyhdr}       
\usepackage{graphicx}       
\graphicspath{{media/}}     
\usepackage{natbib}
\setcitestyle{authoryear} 
\usepackage[figuresright]{rotating}
\usepackage{amsmath}
\usepackage{amsfonts}
\newtheorem{theorem}{Theorem}
\usepackage[T1]{fontenc}
\usepackage{newpxmath}
\usepackage{hyperref}
\hypersetup{
    colorlinks,
    linkcolor={blue},
    citecolor={blue},
    urlcolor={black}
}

\usepackage{lscape}
\usepackage{longtable}
\usepackage{multirow}
\usepackage{threeparttable}
  
\title{Improve the Precision of Area Under the Curve Estimation for Recurrent Events Through Covariate Adjustment}

\author{
  Jiren Sun \\
  Department of Biostatistics and Medical Informatics \\
  University of Wisconsin-Madison \\
  Madison, WI, 53726 \\
\texttt{jiren.sun@wisc.edu} \\
   \And
  Tuo Wang \\
  Global Statistical Science \\
  Eli Lilly and Company \\
  Indianapolis, IN, 46285 \\
  \texttt{tuo.wang@lilly.com} \\
   \AND
   Yanyao Yi \\
  Global Statistical Science \\
  Eli Lilly and Company \\
  Indianapolis, IN, 46285 \\
    \texttt{yi\_yanyao@lilly.com} \\
   \And
   Ting Ye \\
   Department of Biostatistics \\
  University of Washington \\
  Seattle, WA, 98195 \\
    \texttt{tingye1@uw.edu} \\
   \And
   Jun Shao \\
   Department of Statistics \\
   University of Wisconsin-Madison \\
   Madison, WI, 53706 \\
     \texttt{shao@stat.wisc.edu} \\
      \And
   Yu Du \\
  Global Statistical Science \\
  Eli Lilly and Company \\
  Indianapolis, IN, 46285 \\
   \texttt{du\_yu@lilly.com} \\
}

\begin{document}
\maketitle

\begin{abstract}
The area under the curve (AUC) of the mean cumulative function (MCF) has recently been introduced as a novel estimand for evaluating treatment effects in recurrent event settings, offering an alternative to the commonly used Lin-Wei-Yang-Ying (LWYY) model. The AUC of the MCF provides a clinically interpretable summary measure that captures the overall burden of disease progression, regardless of whether the proportionality assumption holds. To improve the precision of the AUC estimation while preserving its unconditional interpretability, we propose a nonparametric covariate adjustment approach. This approach guarantees efficiency gain compared to unadjusted analysis, as demonstrated by theoretical asymptotic distributions, and is universally applicable to various randomization schemes, including both simple and covariate-adaptive designs. Extensive simulations across different scenarios further support its advantage in increasing statistical power. Our findings highlight the importance of covariate adjustment for the analysis of AUC in recurrent event settings, offering practical guidance for its application in randomized clinical trials.
\end{abstract}

\keywords{Area Under the Curve \and Lin-Wei-Yang-Ying (LWYY) Model \and Mean Cumulative Function \and Nonparametric Covariate Adjustment \and Recurrent Events}

\section{Introduction}

In recent years, recurrent event analysis has gained traction in clinical studies, particularly in chronic conditions such as multiple hospitalizations for heart failure, recurrent relapses in multiple sclerosis, or frequent exacerbation in asthma \citep{solomon2024finerenone, rogers2014analysing, cohen2011multiple, keene2007analysis}. A widely adopted approach for analyzing recurrent event data in clinical trials is the Lin-Wei-Yang-Ying (LWYY) model \citep{lin2000semiparametric}. The LWYY model assumes proportionality between the mean cumulative functions (MCFs) of two arms, implying a constant rate ratio over time. This allows treatment effects to be evaluated through the ratio of the MCFs. In practice, this assumption might be violated, particularly in complex scenarios where treatment effects are time-varying. For example, the recent FINEARTS-HF study demonstrated time-dependent variation in rate ratio estimates \citep{vaduganathan2024time}. While the LWYY model offers some robustness to model misspecification \citep{lin1989robust}---and its constant estimate can be interpreted as a weighted average of time-varying effects---this weighted average may lack clear clinical interpretability in some contexts, in which more flexible estimands that do not assume proportionality may be preferable.

The recent ICH E9(R1) addendum also emphasizes the importance of a robust definition quantifying the treatment effects in the settings of recurrent events with minimal assumptions. In this context, the area under the curve (AUC)-based approach for recurrent events proposed by \citet{claggett2022quantifying} has gained considerable attention, offering a clinically meaningful summary of the cumulative burden of recurrent events without assuming proportionality \citep{gregson2023recurrent}. By calculating the area under the MCF for each treatment arm, one can derive a summary measure that reflects the overall mean burden of events experienced by patients over the follow-up period. A larger AUC signifies a greater cumulative burden, and treatment comparisons based on the difference or ratio of AUCs allow for intuitive interpretation in terms of absolute or relative reductions in disease burden. AUCs can be consistently estimated by the Ghosh-Lin estimator irrespective of whether the proportionality assumption holds, making it a robust and interpretable estimand for analyzing recurrent event data \citep{ghosh2000nonparametric}.

Incorporating prognostic baseline covariates into the design and analysis of clinical trial data can improve the efficiency of data utilization, leading to more precise estimation and demonstration of treatment effects \citep{FDA2023, EMA2015}. There are several statistical considerations for better practice in covariate adjustment. First, covariate adjustment should enhance the precision of treatment effect estimation without altering the definition of the estimand. Furthermore, \cite{FDA2023} emphasized that the method used should allow for valid inference under approximately the same minimal statistical assumptions as those required for unadjusted estimation in a randomized trial, thereby avoiding reliance on additional assumptions. One potential approach to covariate adjustment for the AUC is a regression-based method. This approach shares a similar model structure with the LWYY model but replaces the rate function in the LWYY model with the AUC. However, the regression-based approach has notable limitations. First, when covariates are included in the model, the resulting estimates become non-collapsible \citep{hauck1998should}, targeting conditional rather than unconditional treatment effects, which might alter the estimand. Additionally, regression models impose restrictive assumptions, including proportionality, correct specification of covariate functional forms, and the exponential link function. Any of these assumptions may be violated in practice. If the regression model is misspecified, the efficiency gains from covariate adjustment cannot be guaranteed \citep{kong1997robust}.

To address these challenges, we propose a framework for covariate adjustment in the AUC-based summary for recurrent events, extending the methods developed by \cite{ye2022inference, ye2023toward, ye2024covariate}. The primary objective is to enhance the precision of AUC-based summaries while addressing the limitations associated with the regression-based approach and other semiparametric covariate adjustment approaches. Importantly, the covariate-adjusted AUC-based summary preserves the estimand, targeting the unconditional treatment effect. The proposed approach is robust to model misspecification, ensuring that a consistent and valid estimator is obtained regardless of model correctness or the proportionality assumption. This approach guarantees asymptotic efficiency gains over the unadjusted method. Additionally, the approach is universally applicable to simple and covariate-adaptive randomization schemes, achieving optimal efficiency gains, unlike other covariate adjustment approaches that are only applicable to simple randomization \citep{lu2008improving}.

The remainder of the paper is organized as follows. Section \ref{sec:auc} covers the covariate adjustment approach for the AUC estimand for recurrent events. Section \ref{sec:simulations} presents simulation studies that demonstrate the advantage of the covariate-adjusted AUC-based summary in increasing statistical power, and Section \ref{sec:examples} applies the method to real data examples. Section \ref{sec:discussion} concludes with a discussion and future directions.

\section{Area Under the Mean Cumulative Function}\label{sec:auc}

The area under the MCF was introduced \citep{claggett2022quantifying} to address the question: What is the cumulative burden of recurrent events over a fixed follow-up period? It quantifies this burden by capturing both the frequency and timing of recurrent events (e.g., repeated hospitalizations). Conceptually, each occurrence of a recurrent event increases the total time a subject spends in the undesirable condition, contributing to their cumulative burden, which is measured by event-time. The size of this contribution depends on timing: the earlier the event occurs, the more event-time it adds. The MCF tracks the average accumulation of such events across the study population. Therefore, the area under the MCF summarizes the mean total burden due to recurrent events, with a higher AUC indicating greater overall burden. This quantity provides a robust, model-free method for comparing treatment groups in terms of cumulative burden, with treatment effects expressed through differences or ratios of AUCs to reflect absolute or relative reductions in recurrent event burden. In what follows, we formally define the AUC estimand and describe how it can be estimated from censored recurrent event data.

\subsection{Data and estimand}
Let $D$ represent the time to a terminal event and write $N^D(t) = \mathbb{I}(D \leq t)$ as the counting process for $D$. Let $N^*(t)$ be the number of recurrent events up to time $t$. Let $C$ denote the independent censoring time. An observation consists of $\{ N(\cdot), T, \delta \}$, where 
$ N(t) = N^{*}(t \wedge C)$, 
$T = C \wedge D$,  $\delta=\mathbb{I}(D \leq C)$, and $a \wedge b =\hbox{min}(a, b)$. The observed data set $\mathcal{D}=\{ N_{i}(\cdot), T_{i}, \delta_{i} \}$, for $i=1, \dots, n$, consists of independent realizations of $\{ N(\cdot), T, \delta \}$.

Let $S(u)=\mathbb{P}(D \geq u)$ be the survival function, and $dR(u)=\mathbb{E}\{ dN^{*}(u) \mid T \geq u \}=\mathbb{E}\{ dN^{*}(u) \mid D \geq u \}$, where $dN^{*}(t)=\lim_{\Delta t \to 0} \{ N^{*}(t+\Delta t^{-})-N^{*}(t^{-})\}$. Then, the mean cumulative function (MCF) is defined as 
\begin{eqnarray*}
    \mu(t) = \mathbb{E} \{ N^{*}(t) \} = \int_{0}^{t} S(u) dR(u). 
\end{eqnarray*}
The area under the MCF over $[0, \tau]$ is 
\begin{eqnarray*}
    U(\tau) = \int_{0}^{\tau} \mu(t)dt = \int_{0}^{\tau}  \left\{ \int_{0}^{t} S(u) dR(u)  \right\} dt = \int_{0}^{\tau} \int_{0}^{\tau} \mathbb{I} (u \leq t) S(u) dR(u) dt = \int_{0}^{\tau} (\tau - u) S(u) dR(u).
\end{eqnarray*}

The estimand $U(\tau)$ can be interpreted as the mean total event-free time lost from recurrent events over the interval $[0, \tau]$, which captures the cumulative burden of recurrent events in a clinically interpretable way. $U(\tau)$ is a natural extension of the restricted mean survival time (RMST) from time-to-first event settings to the recurrent event setting \citep{kim2017restricted}. Specifically, in the time-to-first event setting, the MCF can be considered as the expected value of the counting process for the death event at time $t$, which is equivalent to the probability of experiencing death by time $t$---that is, 1 minus the survival probability at time $t$. As a result, in the time-to-first event setting, the area under the MCF over the interval $[0, \tau]$ equals $\tau$ minus the RMST over $[0, \tau]$.

Denote $U_0(\tau)$ and $U_1(\tau)$ as the area under the MCF for placebo and treatment arms. The treatment effect can be quantified as either the difference, $\Delta U(\tau)=U_1(\tau) - U_0(\tau)$, or the ratio, $U_1(\tau)/U_0(\tau)$. For convenience in later derivations, we define $R(\tau)= \hbox{log} \frac{U_1(\tau)}{ U_0(\tau)}$, which represents the log-transformed ratio of the areas under the MCF. Similar to how the RMST is estimated based on the Kaplan-Meier estimator, the AUC under each treatment group can be nonparametrically estimated by plugging in the Ghosh-Lin estimator \citep{ghosh2000nonparametric}. An open-source implementation of this method is available at https://github.com/zrmacc/MCC.

An estimand specifies five key attributes: the treatment (the specific intervention being evaluated), population (the group of subjects to which the treatment effect applies), variable (the outcome or endpoint being measured), intercurrent events (events occurring after randomization that may affect the interpretation or occurrence of the outcome), and the population-level summary (how the variable is summarized to compare treatment conditions). The five attributes of our estimand are defined in Table \ref{tab:estimand}.

\subsection{Nonparametric Covariate Adjustment}\label{sec_covadj}

We extend the nonparametric covariate adjustment approach introduced by \cite{ye2022inference, ye2023toward, ye2024covariate} to the AUC-based summary. The key concept involves linearizing the test statistic to derive an outcome for each patient, followed by the application of generalized regression adjustment or augmentation to these derived outcomes \citep{zhang2008improving, tsiatis2008covariate}. In this section, we focus on covariate adjustment for the ratio of AUCs. The corresponding method for the difference of AUCs follows analogously and is provided in Section A of the Supplementary Materials.

\subsubsection{Preparation}
Let $\hat \mu(t)$ and $\hat U(\tau)$ denote estimates of $\mu(t)$ and $U(\tau)$, obtained by substituting $S(u)$ with the Kaplan-Meier estimator $\hat S(u)$ and $dR(u)$ with the Nelson-Aalen estimator $d\hat R(u)= \sum_{i=1}^{n} dN_{i}(u) \big /\sum_{i=1}^{n} \mathbb{I}(T_i \geq u)$, where $ N_i(t) = N_i^{*}(t \wedge C_i)$ and $dN_i(t)=\lim_{\Delta t \to 0} \{ N_i(t+\Delta t^{-})-N_i(t^{-})\}$.

Define $M_i(t)=N_{i}(t) - \int_{0}^{t} \mathbb{I}(T_i \geq u) dR(u) $ as the recurrent event martingale and $M_i^{D}(t)=N_{i}^{D}(t) - \int_{0}^{t} \mathbb{I}(T_i \geq u) dA^{D}(u) $ as the terminal event martingale, where $dA^{D}(u) = \mathbb{E}\{ dN^{D}(u) \mid T \geq u \}$, and $dN^{D}(t)=\lim_{\Delta t \to 0} \{ N^D(t+\Delta t^{-})-N^D(t^{-})\}$, with $N^D_i(t)$ and $dN^{D}_i(t)$ denoting independent realizations of $N^D(t)$ and $dN^{D}(t)$, respectively. Following the line from \cite{zrmacc_mean_2020} the influence function expansion is
\begin{eqnarray*}
    \sqrt{n} \{ \hat U(\tau) - U(\tau)  \} = \frac{1}{\sqrt{n}} \sum_{i=1}^{n} \psi_{i}(\tau) +o_p(1) \label{infl_expansion}
\end{eqnarray*}
where $\psi_{i}(\tau) = P_i - Q_i$, $P_i = \int_{0}^{\tau} \frac{(\tau -u)S(u) }{\mathbb{P}(T \geq u)} dM_{i}(u)$, $Q_i = \int_{0}^{\tau} \frac{ \int_{u}^{\tau} (\tau - s)d\mu(s) }{\mathbb{P}(T \geq u)} dM_{i}^{D}(u)$, and $o_p(1)$ denotes a term tending to 0 in probability as the sample size $n \to \infty$. 

The estimates $\hat \psi_{i}(\tau)$, $\hat P_i$, and $\hat Q_i$ are obtained by replacing $S(u)$ with $\hat S(u)$, $\mu(s)$ with $\hat \mu(s)$, and estimating $\mathbb{P}(T \geq u)$ as $\sum_{i=1}^{n} \mathbb{I}(T_i \geq u)/n$. The martingales are estimated by:
$d\hat M_{i}(u) = dN_{i}(u) - \mathbb{I}(T_i \geq u) d\hat R(u)$, and $d\hat M_{i}^{D}(u) = dN_{i}^{D}(u) - \mathbb{I}(T_i \geq u) d \hat A^{D}(u)$, with $d \hat A^{D}(u) = \sum_{i=1}^{n} dN^{D}_{i}(u)\big/\sum_{i=1}^{n} \mathbb{I}(T_i \geq u)$.

\subsubsection{Covariate-Adjusted AUCs} 
Let $\hat R(\tau)= \hbox{log} \frac{\hat U_1(\tau)}{\hat U_0(\tau)}$ and $\hat{\mathcal{U}}_{L, R} = \frac{n_0 n_1}{n} \left\{ \hat R(\tau) - R(\tau) \right\} $, where $n_0$ and $n_1$ represent the sample sizes for the placebo and treatment arms, respectively. The linearization of $\hat{\mathcal{U}}_{L, R}$ is as follows:
\begin{eqnarray*}
    \hat{\mathcal{U}}_{L, R} = \frac{1}{n} \sum_{i=1}^{n} \left\{ I_{i} (\hat P^{R}_{i} - \hat Q^{R}_{i}) - (1-I_i) (\hat P^{R}_{i} - \hat Q^{R}_{i}) \right\} +o_p(1)
\end{eqnarray*}
where $I_{i}=1$ if subject $i$ is in the treatment arm and 0 otherwise, $\hat P^{R}_{i} = \left\{ \frac{n_0}{\hat U_{1}(\tau)}  \right\}^{I_i}  \left\{ \frac{n_1}{\hat U_{0}(\tau)} \right\}^{1-I_i}   \hat P_i$ and $\hat Q^{R}_{i} = \left\{ \frac{n_0}{\hat U_{1}(\tau)}  \right\}^{I_i}  \left\{ \frac{n_1}{\hat U_{0}(\tau)} \right\}^{1-I_i}  \hat Q_i$. The unadjusted test statistic on the ratio of AUCs is $\mathcal{T}_{L,R} = \sqrt{n}  \hat{\mathcal{U}}_{L, R}/\hat{\sigma}_{L, R}$, where 
$$\hat{\sigma}_{L, R}^{2} = \frac{1}{n}  \left[ \frac{n_0^2 n_1}{\hat U_{1}(\tau)^{2}}  \mathbb{E}\left\{ \hat \psi_{1}(\tau)^{T} \hat \psi_{1}(\tau) \right\} + \frac{n_1^2 n_0}{\hat U_{0}(\tau)^{2}}  \mathbb{E}\left\{ \hat \psi_{0}(\tau)^{T} \hat \psi_{0}(\tau)\right\} \right].$$ Here, $\hat \psi_{1}(\tau)$ and $\hat \psi_{0}(\tau)$ consist of $\hat \psi_{i}(\tau)$ from subjects in the treatment and placebo arms, respectively. Derivations of $\hat{\mathcal{U}}_{L,R}$ and the asymptotic properties of $\mathcal{T}_{L,R}$ are provided in detail in Section B of the Supplementary Materials.

Let $\boldsymbol{X}^{*}$ be a $p$-dimensional vector of baseline covariates. Define $X_{i}=X_{i}^{*}-\bar X$, where $X_{i}^{*}$ is the $i$th column of $\boldsymbol{X}^{*}$, and $\bar{X}$ is the sample mean of all the $X_{i}^{*}$, ensuring $\mathbb{E}(X_i)=0$. Following \cite{ye2024covariate}, we treat $\hat P^{R}_{i}$ and $\hat Q^{R}_{i}$ as outcomes and apply the generalized regression adjustment or augmentation \citep{tsiatis2008covariate}. This results in the covariate-adjusted $\hat{\mathcal{U}}_{CL,R}$:
\begin{eqnarray*}
    \hat{\mathcal{U}}_{CL,R} = \frac{1}{n} \sum_{i=1}^{n} \Bigg[ I_{i} \left\{ (\hat P^{R}_{i} - X_{i}^{T} \hat \beta_{1,P}^{R}) 
    - (\hat Q^{R}_{i} - X_{i}^{T} \hat \beta_{1,Q}^{R}) \right\}  - (1-I_i) \left\{ (\hat P^{R}_{i} - X_{i}^{T} \hat \beta_{0,P}^{R}) 
    - (\hat Q^{R}_{i} - X_{i}^{T} \hat \beta_{0,Q}^{R}) \right\} \Bigg]  = \hat{\mathcal{U}}_{L,R} - \hat{\mathcal{A}_{R}} 
\end{eqnarray*}
where 
\begin{eqnarray*}
    \hat{\mathcal{A}_{R}} &=& \frac{1}{n} \sum_{i=1}^{n} \left\{ I_i  X_{i}^{T}  \left( \hat \beta_{1,P}^{R} - \hat \beta_{1,Q}^{R}  \right) \right. - \left. (1-I_i)  X_{i}^{T}  \left( \hat \beta_{0,P}^{R} - \hat \beta_{0,Q}^{R}  \right) \right\}, \\
    \hat \beta_{j,P}^{R} &=& \left( \sum_{i:I_i=j} X_i X_{i}^{T} \right)^{-1} \sum_{i:I_i=j} X_{i} \hat P^{R}_{i}, \\
    \hat \beta_{j,Q}^{R} &=& \left( \sum_{i:I_i=j} X_i X_{i}^{T} \right)^{-1} \sum_{i:I_i=j} X_{i} \hat Q^{R}_{i}.
\end{eqnarray*}

Let $\hat R^{(a)}(\tau)$ denote the adjusted $ \hat R(\tau)$, and is given by $\hat R^{(a)}(\tau)=\hat R(\tau) - \frac{n}{n_0 n_1} \hat{\mathcal{A}_{R}}$. Then, $\hat{\mathcal{U}}_{CL,R}= \frac{n_0 n_1}{n} \left\{ \hat R^{(a)}(\tau) - R(\tau) \right\}$. Define the covariate-adjusted test statistic on the ratio of AUCs as $\mathcal{T}_{CL,R} = \sqrt{n}  \hat{\mathcal{U}}_{CL, R}/\hat{\sigma}_{CL, R}$, where $\hat{\sigma}_{CL, R}^2= \hat{\sigma}_{L, R}^2 - \frac{n_0 n_1}{n^2} (\hat \beta_{1,P}^{R} - \hat \beta_{1,Q}^{R} + \hat \beta_{0,P}^{R} - \hat \beta_{0,Q}^{R})^{T}  \sum_{X}  (\hat \beta_{1,P}^{R} - \hat \beta_{1,Q}^{R} + \hat \beta_{0,P}^{R} - \hat \beta_{0,Q}^{R}) $. The theorem below establishes the asymptotic properties of $\hat{\mathcal{U}}_{CL,R}$ and $\mathcal{T}_{CL, R}$. The technical proofs are included in Section C of the Supplementary Materials.
\begin{theorem}\label{theoadjratio}
    The following results hold under both simple randomization and covariate-adaptive randomization schemes:
    \\
    Under the $H_{0,R}: R(\tau)=0$ or the alternative hypothesis, $\sqrt{n}  \hat{\mathcal{U}}_{CL,R}\stackrel{d}{\to} N\left(0, \sigma_{CL, R}^{2}\right) $, $\hat{\sigma}_{CL, R}^{2} \stackrel{p}{\to} \sigma_{CL, R}^{2}$, and $\mathcal{T}_{CL, R} \stackrel{d}{\to} N(0, 1) $, where $\sigma_{CL, R}^2= \sigma_{L, R}^2 - \frac{n_0 n_1}{n^2} ( \beta_{1,P}^{R} - \beta_{1,Q}^{R} + \beta_{0,P}^{R} - \beta_{0,Q}^{R})^{T}   \sum_{X}  ( \beta_{1,P}^{R} - \beta_{1,Q}^{R} + \beta_{0,P}^{R} - \beta_{0,Q}^{R}) $. 
\end{theorem}

$H_{0,R}$ is rejected when $|\mathcal{T}_{CL,R}| > z_{\alpha/2}$ and the adjusted $100 (1-\alpha) \%$ CI for $R(\tau)$ is $\big(\hbox{exp}\big[\hat R^{(a)}(\tau) - z_{\alpha/2}  SE\left\{ \hat R^{(a)}(\tau) \right\} \big], \hbox{exp}\big[\hat R^{(a)}(\tau) + z_{\alpha/2}  SE\left\{ \hat R^{(a)}(\tau) \right\} \big]\big)$, where $SE \left\{ \hat R^{(a)}(\tau)\right\}= \left( \frac{n}{n_0^2 n_1^2}  \hat{\sigma}_{CL, R}^2 \right)^{1/2}$. 

As proved by \cite{ye2024covariate}, $\sigma_{CL, R}^2 \leq \sigma_{L, R}^2$, with strict inequality holding unless either (i) the covariates $X_{i}^{*}$ are uncorrelated with $P_{i}^{R}$ and $Q_{i}^{R}$, or (ii) the covariates not used in the randomization are uncorrelated with $P_{i}^{R}$ and $Q_{i}^{R}$, given the covariates used in the randomization. Therefore, the adjusted $\mathcal{T}_{CL, R}$ provides a guaranteed efficiency gain over the unadjusted $\mathcal{T}_{L, R}$.

\section{Simulations}\label{sec:simulations}
\subsection{Data Generation}
Three independent baseline covariates $\boldsymbol{X}^{*} = (X_1^{*}, X_2^{*}, X_3^{*})$ are considered. Specifically, $X_1^{*}$ follows a Bernoulli distribution with probability of 0.5 while $X_2^{*}$ and $X_3^{*}$ independently follow a normal distribution with mean of 0 and standard deviation of 2.

Two randomization schemes with equal allocation are considered. The first scheme is simple randomization, where treatments are assigned to subjects completely at random. This approach may result in imbalances across the treatment arms concerning $(X_1^{*}, X_2^{*}, X_3^{*})$. To mitigate this, covariate-adaptive randomization, which aims to balance treatment arms with respect to baseline prognostic factors, is commonly used. A popular method is stratified permuted block (SPB) randomization \citep{ZELEN1974365}. However, practical constraints may prevent the inclusion of all baseline prognostic factors in covariate-adaptive randomization. In this simulation study, $X_1^{*}$ and $X_2^{*}$ are included in the SPB randomization, while all three covariates are used for adjustment during the analysis stage. In SPB randomization, subjects are categorized into strata based on $X_1^{*}$ and $X_2^{*}$, where $X_2^{*}$ is categorized into four quantiles. Within each stratum, subjects are further grouped into blocks of size 4, and permuted block randomization is applied within these blocks.

There are a total of 2,000 subjects. For each subject $i$, the death time $D_i=5/365+\omega_{i}$, where $\omega_{i}$ follows an exponential distribution with the mean as $\left\{ 0.05  \hbox{exp}( X_{i}^{*T} \xi) \right\}^{-1}$, where $\xi=(0.1, 0.1, 0.1)^{T}$. The maximum follow-up time for each subject is 2 years. The independent censoring $C_i$ is uniformly distributed on $(1,2)$. 

Recurrent event data are generated under the following five cases. Let $\theta$ denotes a scalar parameter controlling the magnitude of treatment effects, with $\eta = (0.2, 0.2, 0.2)^{T}$, and $\rho_{0}(t) = 0.3t$. $\tau$ is set as 2 when calculating AUCs. 

\begin{itemize}
    \item Case 1: Recurrent events are generated from the conditional intensity function that follows an Anderson-Gill model: $\rho^{(1)}_{j}(t \mid \boldsymbol{X}^{*}) = \rho_{0}(t)  \hbox{exp}(\theta j + \boldsymbol{X}^{*T} \eta )$ for $j=0,1$. The treatment effect characterized by $\theta$ remains constant over time. 
    \item Case 2: Recurrent events are generated using the conditional intensity function: $\rho^{(2)}_{j}(t \mid \boldsymbol{X}^{*}) = \rho_{0}(t)  \hbox{exp}\{\theta^{(1)}(t)  j + \boldsymbol{X}^{*T} \eta \}$ for $j=0,1$, where $\theta^{(1)}(t) = -0.25 \theta t^2+\theta t$. The treatment effect is characterized by a time-varying $\theta^{(1)}(t)$, which moves from 0 at $t=0$ to $\theta$ at $t=2$. This case features a late treatment effect scenario, where the treatment effect gradually increases and reaches its full extent by the end.
    \item Case 3: Recurrent events are generated from the conditional intensity function: $\rho^{(3)}_{j}(t \mid \boldsymbol{X}^{*}) = \rho_{0}(t)  \hbox{exp}\{\theta^{(2)}(t)  j + \boldsymbol{X}^{*T} \eta \}$ for $j=0,1$, where $\theta^{(2)}(t) = -\theta t^2+2\theta t$. $\theta(t)$ moves from 0 at $t=0$, reaches a peak of $\theta$ at $t=1$, and backs to 0 at $t=2$. This case exhibits a peak treatment effect scenario, with the effect being strongest at a specific point and weaker before and after that peak.
    \item Case 4: Recurrent events are generated from the conditional intensity function: $\rho^{(4)}_{j}(t \mid \boldsymbol{X}^{*}) = \rho_{0}(t)  \hbox{exp}\{\theta^{(3)}(t)  j + \boldsymbol{X}^{*T} \eta \}$ for $j=0,1$, where $\theta^{(3)}(t) = -0.25\theta t^2+\theta$. $\theta(t)$ moves from $\theta$ at $t=0$ to 0 at $t=2$. This case presents an early treatment effect scenario, where the effect is immediate at the beginning and then diminishes over time.
    \item Case 5: Let $\Delta V_{ij}$ be the gap time between the $(j-1)$th and $j$th recurrent event for subject $i$. Define $\Delta V_{ij} = \hbox{exp}(-\theta  I_i + X_{i}^{*T} \eta - 0.7) + \epsilon_{j}$, where $\epsilon_{j}$ is an independent random variable following an exponential distribution with mean as 0.25. The ordered $j$th recurrent event time for subject $i$ is then $\sum_{k=1}^{j} \Delta V_{ik}$, for $\sum_{k=1}^{j} \Delta V_{ik} < T_i$, with the $0$th recurrent event time being 0. Case 5 differs from the preceding cases in that the treatment affects recurrent events by reducing the gap time between them. 
\end{itemize}

\subsection{Results}

The mean follow-up across the five cases and two randomization schemes is 1.44 years, with 7.8\% of subjects experiencing mortality during the study period. The average number of recurrent events per subject is the same between the two randomization schemes but varies slightly between cases. For the placebo arm, the average number of recurrent events per subject is 0.68 for Cases 1 through 4 and 0.81 for Case 5. In the treatment arm, with $\theta=-0.16$, the corresponding averages are 0.58, 0.59, 0.61, 0.63, and 0.72 for Cases 1 through 5, respectively. When $\theta=-0.32$, the averages decrease further to 0.49, 0.51, 0.55, 0.58, and 0.62 for Cases 1 through 5, respectively.

\begin{figure}
    \centering
    \includegraphics[width=\textwidth]{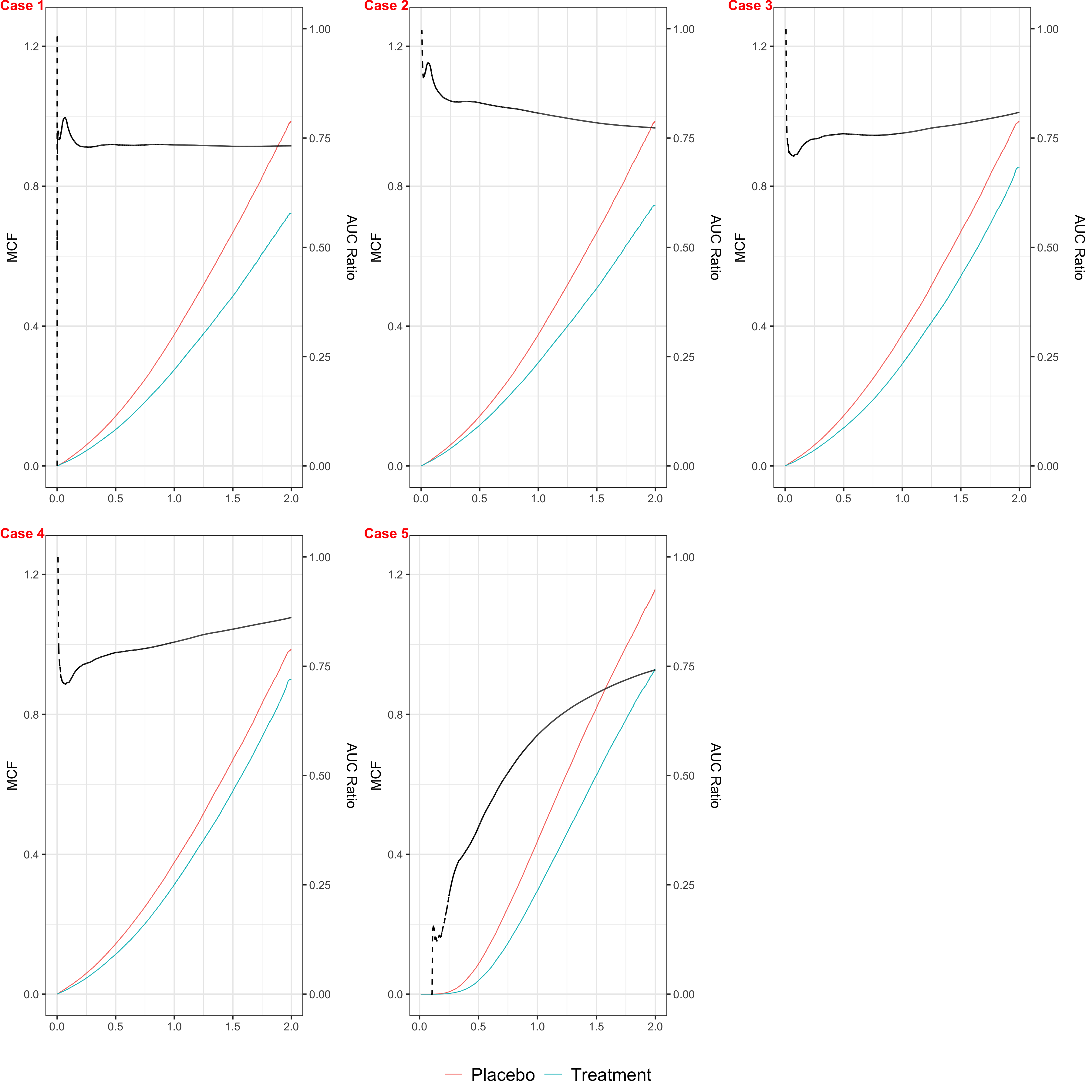}
    \caption{Mean cumulative functions for different simulation cases using the simple randomization scheme. Each case is based on data from a single simulation with a total sample size of 100,000 (allocated in a 1:1 ratio). The scalar $\theta$ is set to -0.32. The ratio $\hat U_1(t)/\hat U_0(t)$ at each $t$ is represented by the black curve. }
    \label{mcf}
\end{figure}

Figure \ref{mcf} shows the MCFs for the five cases with the simple randomization scheme, using a total sample size of 100,000 with equal allocation to generate smoother curves. $\theta$ is set to -0.32. The treatment effect is constant in Case 1 but not in Cases 2 to 5. However, it is difficult to detect the time-varying treatment effects from MCFs. To better view the treatment effect over time, the black curve in Figure \ref{mcf} represents $\hat{U}_1(t) / \hat{U}_0(t)$ over time. Despite some instabilities at earlier times, $\hat{U}_1(t) / \hat{U}_0(t)$ is constant over time in Case 1. In Case 2, $\hat{U}_1(t) / \hat{U}_0(t)$ decreases over time. In Case 3, the trend of $\hat{U}_1(t) / \hat{U}_0(t)$ is not very clear before year 1, but after year 1, $\hat{U}_1(t) / \hat{U}_0(t)$ shows an obvious upward trend. In Case 4, $\hat{U}_1(t) / \hat{U}_0(t)$ increases over time, and in Case 5, $\hat{U}_1(t) / \hat{U}_0(t)$ remarkably increases early on and becomes stable near the tail.

Table \ref{tab:auc_results} presents simulation results for the difference and ratio of AUCs from both unadjusted and covariate-adjusted analyses across various scenarios and randomization schemes, based on 5,000 simulations. For practical purposes, we use the Monte Carlo mean of the 5,000 estimated unadjusted differences and ratios of AUCs---denoted as $\Delta \hat U(\tau)$ and $\hat R(\tau)$---as empirical approximations of the true values $\Delta U(\tau)$ and $R(\tau)$, and label them as ``Est.'' ``Bias'' refers to the difference between the Monte Carlo mean of the 5,000 adjusted estimates ($\Delta \hat U^{(a)}(\tau)$ and $\hat R^{(a)}(\tau)$) and the corresponding ``Est.'' As expected, the ``Bias'' is consistently zero across all scenarios, empirically confirming that the proposed covariate adjustment procedure preserves the estimand. ``Mean'' and ``Median'' denote the mean and median of the 5,000 estimated standard errors (SEs) for each point estimate---$\Delta \hat U(\tau)$, $\Delta \hat U^{(a)}(\tau)$, $\hat R(\tau)$, and $\hat R^{(a)}(\tau)$---while ``MC'' represents the Monte Carlo standard deviation of the corresponding estimates. Coverage probability (CP) refers to the likelihood that the 95\% confidence interval (CI) contains the respective ``Est.'' For the adjusted analyses, ``Mean,'' ``Median,'' and ``MC'' are nearly identical across all settings, and the CP remains close to the nominal level. In contrast, under the SPB randomization scheme, the unadjusted analysis yields ``Mean'' and ``Median'' values consistently larger than ``MC,'' with CP exceeding the nominal level---suggesting that unadjusted analyses that ignore covariate-adaptive randomization may be conservative. This observation aligns with FDA guidance on covariate adjustment, which recommends adjusting for baseline covariates in the analysis if they were balanced during randomization \citep{FDA2023}. Additional results for a smaller sample size (total of 400 subjects) are provided in Section D.1 of the Supplementary Materials. The results are similar to those presented in Table \ref{tab:auc_results}.

\begin{figure}
    \centering
    \includegraphics[width=\textwidth]{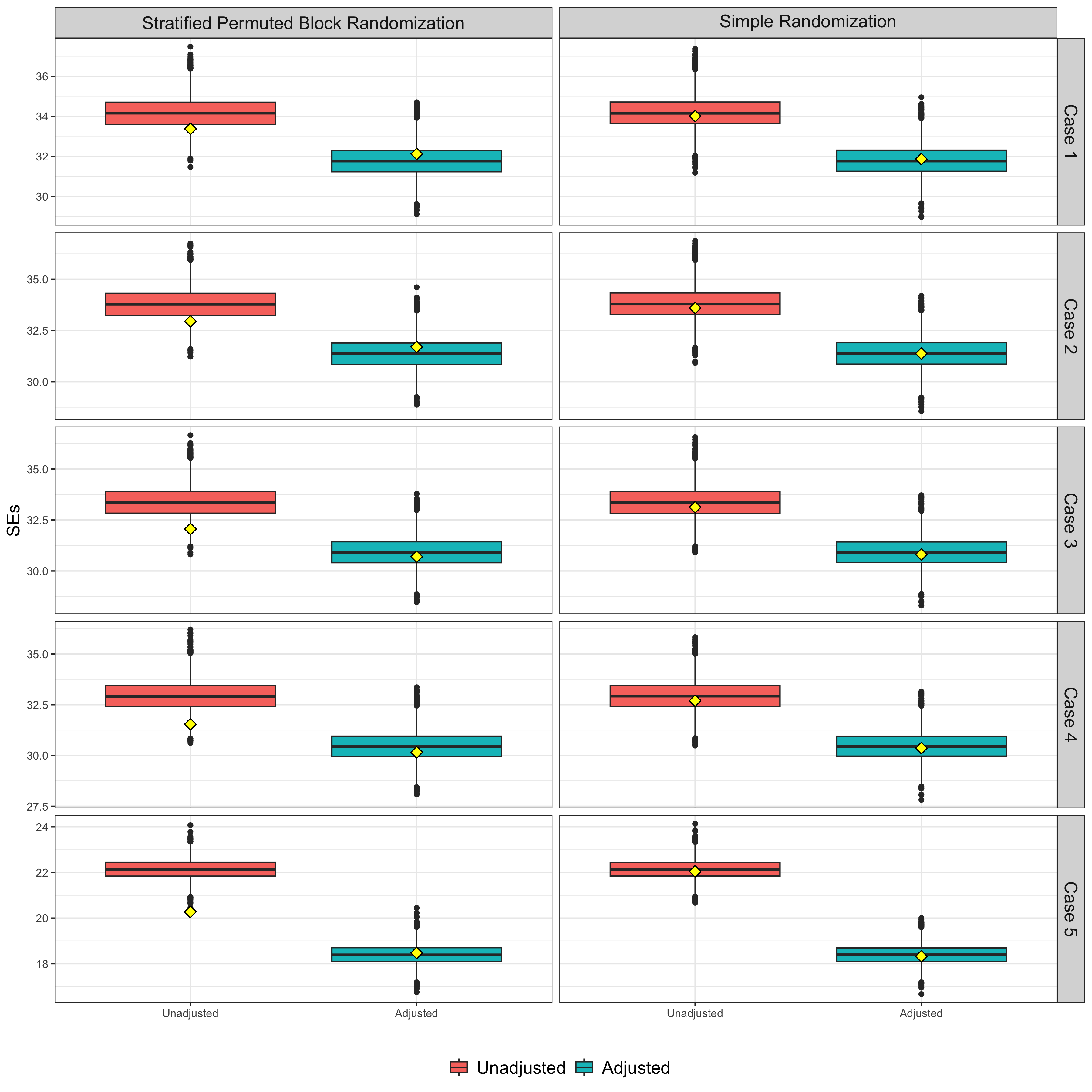}
    \caption{Estimated standard errors of $\sqrt{\hat{\sigma}_{L, R}^{2}/n}$ (unadjusted) and $\sqrt{\hat{\sigma}_{CL, R}^{2}/n}$ (adjusted) from 5,000 simulations under different randomization schemes and cases. Yellow diamonds represent the Monte Carlo standard deviations of 5,000 replicates of $\hat{\mathcal{U}}_{L,R}$ (unadjusted) and $\hat{\mathcal{U}}_{CL,R}$ (adjusted). The scalar $\theta$ is set to -0.32. $\tau$ is set to 2.}
    \label{se}
\end{figure}

To visually demonstrate the validity of our variance formulas and highlight the conservativeness of the unadjusted analysis under covariate-adaptive randomization, Figure \ref{se} displays the distribution of estimated standard errors $\sqrt{\hat{\sigma}_{L, R}^{2}/n}$ (unadjusted) and $\sqrt{\hat{\sigma}_{CL, R}^{2}/n}$ (adjusted) from 5,000 simulations under different randomization schemes and cases, with $\theta=-0.32$ and $\tau=2$. Similar results for $\sqrt{\hat{\sigma}_{L, \Delta}^{2}/n}$ and $\sqrt{\hat{\sigma}_{CL, \Delta}^{2}/n}$ are presented in Section D.2 of the Supplementary Materials. Yellow diamonds represent the Monte Carlo standard deviations of 5,000 $\hat{\mathcal{U}}_{L,R}$ (unadjusted) and $\hat{\mathcal{U}}_{CL,R}$ (adjusted). In the unadjusted analysis, the yellow diamond for simple randomization is higher compared to SPB randomization. This occurs because SPB randomization reduces variability by enforcing balance with respect to $X^{*}_{1}$ and $X^{*}_{2}$. For simple randomization in the unadjusted analysis, as well as for both simple and SPB randomization in the adjusted analysis, the boxplot bodies are thin and include the yellow diamond, despite some outliers. This empirically verifies the asymptotic properties outlined in Theorem \ref{theoadjratio}, suggesting that our variance formulas are accurate and that the covariate adjustment approach is applicable to both simple and covariate-adaptive randomization. Conversely, in SPB randomization, the boxplots from the unadjusted analysis are positioned above the corresponding yellow diamonds, indicating that the estimated standard errors overestimate the actual standard deviations, making the analyses conservative. This observation further supports the notion that standard errors can be overestimated when covariate-adaptive randomization schemes are ignored.

\begin{figure}
    \centering
    \includegraphics[width=\textwidth]{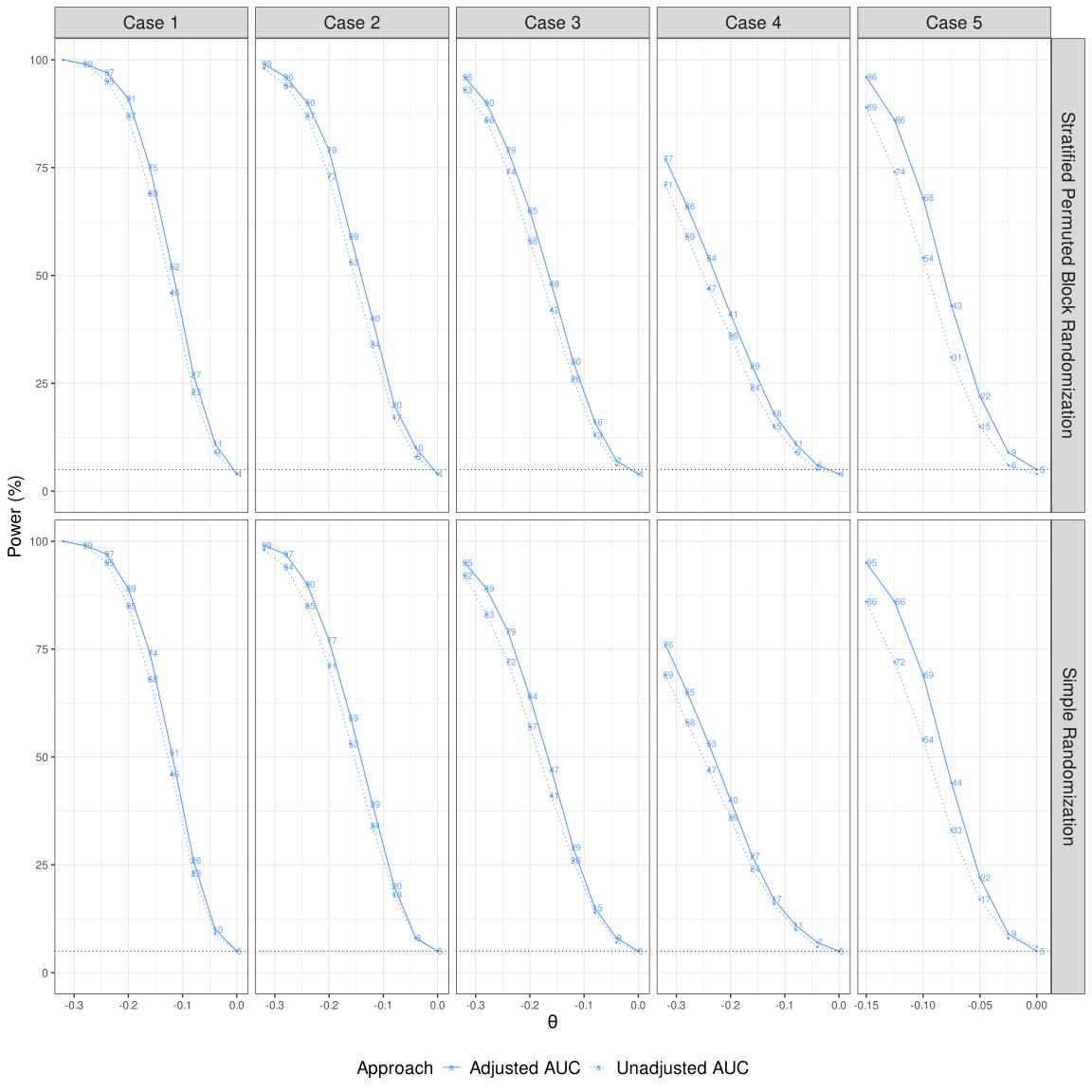}
    \caption{Comparison of power based on 5,000 simulations under different randomization schemes and cases. The dotted horizontal line indicates the significance level, $\alpha = 5\%$. $\tau$ is set to 2.}
    \label{power_auc_only}
\end{figure}

Figure \ref{power_auc_only} compares the power of unadjusted AUC with adjusted AUC based on 5,000 simulations. Tests on the difference and ratio of AUCs show identical power. The figure demonstrates that the adjusted analysis consistently yields higher power across various cases and randomization schemes. This empirically supports the efficiency gain suggested by Theorem~\ref{theoadjratio}. 

\section{Real Examples}\label{sec:examples}

\subsection{HF-ACTION}
The HF-ACTION trial (Heart Failure: A Controlled Trial Investigating Outcomes of Exercise Training) was a multicenter, randomized controlled trial designed to evaluate the effects of structured exercise training on patients with chronic heart failure and reduced ejection fraction \citep{o2009efficacy}. A total of 2,331 participants were randomly assigned to either an exercise training program or a control arm receiving usual care, stratified by center and heart failure etiology. The primary objective was to assess the impact of exercise training on a composite endpoint of all-cause hospitalization and mortality.

Our analysis focuses on a subgroup of 426 non-ischemic patients who had a baseline cardiopulmonary exercise test duration of nine minutes or less. Of these, 205 were assigned to the exercise training arm, and 221 to usual care. Over a median follow-up of approximately 28 months, mortality rates were about 18\% in the exercise training arm compared to 26\% in the usual care arm. The average number of recurrent hospitalizations per patient was 2.2 for the exercise arm and 2.6 for the usual care arm.

\begin{figure}
    \centering
    \includegraphics[width=\textwidth]{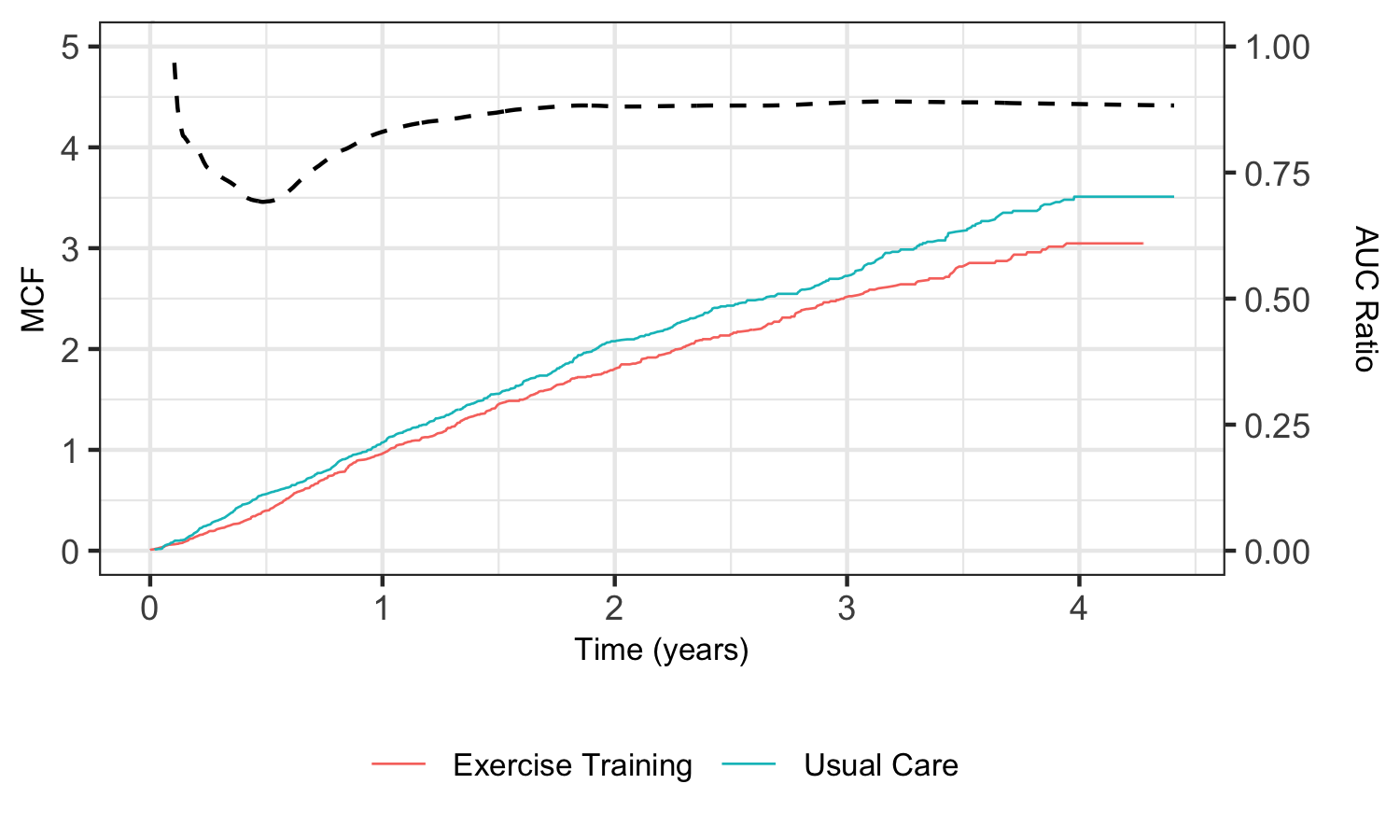}
    \includegraphics[width=\textwidth]{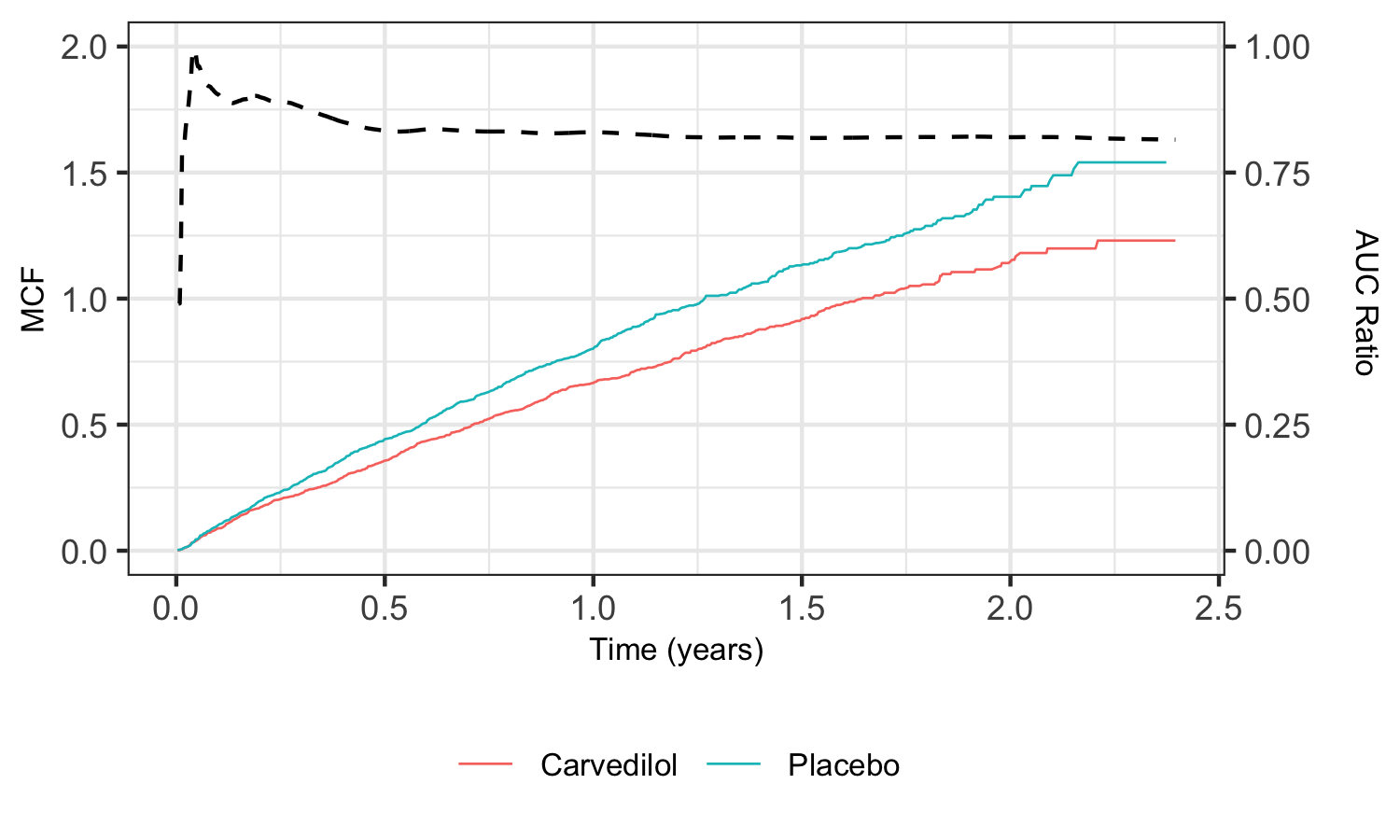}
    \caption{Mean cumulative functions from the HF-ACTION trial (top) and COPERNICUS trial (bottom). The ratio $\hat U_1(t)/\hat U_0(t)$ at each $t$ is represented by the black curve.}
    \label{hfactionmcf}
\end{figure}

The upper panel of Figure \ref{hfactionmcf} illustrates the mean cumulative function for both arms, along with the ratio of AUCs $\hat U_1(t)/\hat U_0(t)$ over time, represented by the black curve. Although there is some instability in the black curve at earlier time points, it remains relatively constant over time. This suggests that the constant treatment effect assumption of the LWYY model may not be significantly violated. However, for the purpose of illustrating our covariate adjustment approach for AUC, we will proceed with an AUC analysis.

In the analysis, we use an age indicator (whether a patient is over 60) as a covariate. Table \ref{hfactiontable} shows the estimated ratio and difference of AUCs between the two arms over the time window $[0,4]$. After adjusting for age, the variance of the log ratio of AUCs decreases from 0.0151 to 0.0147 (a 2.65\% reduction), and the variance of the difference in AUCs decreases from 0.7695 to 0.7526 (a 1.02\% reduction). These reductions suggest that covariate adjustment improves the precision of our estimates.

\subsection{COPERNICUS}
The COPERNICUS trial (Carvedilol Prospective Randomized Cumulative Survival) was a double-blind, placebo-controlled, multicenter study investigating the effect of carvedilol, a nonselective beta-blocker, on morbidity and mortality in patients with severe heart failure \citep{packer2001effect}. The study enrolled 2,289 participants, with 1,156 assigned to carvedilol and 1,133 to placebo. The primary outcome, death from any cause, occurred in 130 patients (11.2\%) in the carvedilol arm compared to 190 patients (16.8\%) in the placebo arm.

There were 700 hospitalizations in the carvedilol arm versus 848 in the placebo arm. The average number of recurrent hospitalizations per patient was 0.606 for the carvedilol arm compared to 0.748 for the placebo arm.

The lower panel of Figure \ref{hfactionmcf} shows the MCF for both arms, along with the ratio of AUCs, $\hat U_1(t)/\hat U_0(t)$, over time (black curve). The treatment effect appears to remain relatively constant over time, making the LWYY model an appropriate choice for this data. However, for illustration purposes, we will proceed with an AUC analysis. Covariates in this analysis include sitting heart rate, history of myocardial infarction, presence of concomitant disease, diabetes, and the use of diuretics, ACE inhibitors, digitalis, amiodarone, and anticoagulants. Table \ref{hfactiontable} provides the estimated ratio and difference of AUCs between the two arms over the time window $[0,2.2]$. After adjusting for covariates, the variance of the log ratio of AUCs decreases from 0.0055 to 0.0052 (a 5.77\% reduction), and the variance of the difference in AUCs decreases from 0.0153 to 0.0146 (a 4.79\% reduction). These reductions further demonstrate that covariate adjustment improves the precision of our estimates.

\subsection{Practical Implications}
In this section, we demonstrate our covariate adjustment approach for AUC using two real-world clinical trials: the HF-ACTION trial and the COPERNICUS trial. For HF-ACTION, we have access only to a subgroup of participants with a single available covariate. The COPERNICUS trial includes all subjects, but some baseline covariates have missing values. While imputation is a viable strategy, we limit our analysis to covariates with complete data for illustrative purposes. Incorporating additional covariates would likely yield even greater efficiency gains.
 
The relative efficiency between two estimators is defined as the ratio of their variances. In clinical trial design, this can also be interpreted as the ratio of sample sizes required to achieve the same statistical power. In the COPERNICUS example, we observe an approximate 5\% reduction in variance, which can be viewed as a potential 5\% reduction in required sample size. A similar efficiency gain was observed in previous covariate adjustment studies \citep{ye2022inference,ye2023toward,ye2024covariate}.

Such efficiency gains are particularly meaningful in the context of modern clinical trials. As standard care continues to improve and event rates decline, larger sample sizes are often necessary to observe a sufficient number of events. For instance, while COPERNICUS (1997--2000) and HF-ACTION (2003--2007) enrolled fewer participants, many contemporary cardiovascular trials now exceed 10,000 subjects. A 5\% reduction in required sample size could correspond to 500 fewer subjects in a 10,000-person trial---offering substantial savings in time, cost, and patient burden.

\section{Discussion}\label{sec:discussion}

Covariate-adaptive randomization is commonly used in clinical trials. Our simulation studies, along with previous research and FDA's guidance, \cite{FDA2023} have demonstrated the necessity of adjusting for covariates balanced during randomization in subsequent analyses. Failure to do so can result in conservative estimates. In this paper, we aimed to improve the precision of AUC estimation with minimal assumptions under both simple and covariate-adaptive randomization, while preserving unconditional interpretations, using the nonparametric covariate adjustment approach proposed by \cite{ye2022inference, ye2023toward, ye2024covariate}.

Theoretical asymptotic distributions demonstrate that our approach achieves efficiency gains, and extensive simulations across various scenarios and real data analysis confirm its effectiveness in improving statistical power. This nonparametric approach offers several advantages. First, this approach is model-free. Although we use simple linear regression to represent $\mathbb{E}(\hat P_{i}^{\Delta})$, $\mathbb{E}(\hat Q_{i}^{\Delta})$, $\mathbb{E}(\hat P_{i}^{R})$, and $\mathbb{E}(\hat Q_{i}^{R})$ during adjustment, the approach is valid regardless of whether the linear regression model correctly specifies the true expected values. In other words, its validity does not depend on parametric modeling assumptions \citep{zhang2008improving}. The choice of linear regression is for implementation convenience and offers good theoretical properties. This model-free characteristic aligns with the ``minimal statistical assumption'' principle advocated in the recent FDA guidance on covariate adjustment \citep{FDA2023}. Second, unlike conventional regression-based approaches for covariate adjustment, which can lose power if the model is misspecified, the nonparametric covariate adjustment guarantees an efficiency gain over the unadjusted test. Third, this approach is universally applicable to both simple and covariate-adaptive randomization schemes, unlike some nonparametric methods that are limited to simple randomization \citep{lu2008improving}. Lastly, unlike non-linear regression models where the distinction between conditional and unconditional estimates is pronounced (known as non-collapsibility), this approach does not alter the unconditional interpretations.

In summary, three key principles underpin better practice in covariate adjustment: (1) preserving unconditional interpretation, (2) requiring minimal statistical assumptions, and (3) being universally applicable to both simple and covariate-adaptive randomization. Nonlinear regression models clearly do not satisfy principles (1) and (2). While some other covariate adjustment approaches satisfy principles (1) and (2), they are only applicable to simple randomization \citep{lu2008improving}. The nonparametric covariate adjustment approach proposed by \cite{ye2022inference, ye2023toward, ye2024covariate} satisfies all three principles and is cited and recommended in the FDA guidance on covariate adjustment \citep{FDA2023}.

While the unconditional treatment effect is often the primary focus in most randomized trials due to its relevance for broad policy recommendations and regulatory authorities \citep{tsiatis2008covariate}, some researchers argue that the conditional treatment effect is more appropriate for primary inferences. For instance, \cite{hauck1998should} advocate for adjusting for important prognostic covariates to achieve the most clinically relevant subject-specific measure of treatment effect. Inferences on the conditional effect can reveal interactions between treatment and patient characteristics, which may have significant implications for treatment use in specific subpopulations \citep{tsiatis2008covariate}. Both unconditional and conditional treatment effects are important for a comprehensive understanding of treatment comparisons. However, this paper does not engage in this debate. Instead, we focus on covariate adjustment in the unconditional effect, aiming to improve the efficiency using minimal assumptions under different randomization schemes.

The AUC method has certain limitations. First, as discussed by \cite{claggett2022quantifying}, it is most appropriate when death rates are low and balanced across treatment arms. This condition is also most appropriate for the LWYY model. If a treatment prolongs survival, the AUC of the treatment arm is likely to be greater than that of the placebo arm, even if the recurrent event rate is lower in the treatment arm. One potential solution is to use the while-alive estimand \citep{schmidli2023estimands, wei2023properties, mao2023nonparametric}. Second, with small sample sizes, the MCF can be unstable near the tail, making the estimate sensitive to the selection of the time window for analysis. The time window should be prespecified in the study protocol but may be chosen empirically at analysis time with clinical justification \citep{tian2020empirical}.

To the best of our knowledge, this is the first paper to provide a robust, unconditional covariate-adjusted inference for the area under the curve estimand in the recurrent event setting, covering covariate-adaptive randomizations. The curve can represent not only the mean cumulative function but also the survival curve. The application of this approach to RMST is straightforward and discussed in Section E of the Supplementary Materials. While not the primary focus of this paper, we also examine the relative power of the LWYY and AUC approaches across various scenarios in Section D.3 of the Supplementary Materials, acknowledging that they test different null hypotheses. Future research on AUC may include sample size calculation and interim analysis.


\bibliography{maintext}  

\clearpage

\clearpage

\begin{longtable}{ l | p{10cm} }
\caption{Description of the area under the mean cumulative function for recurrent events under the estimand framework.\label{tab:estimand}}  \\
\toprule
\textbf{Estimand Attributes} & \textbf{Description} \\
\hline
\endfirsthead
\hline
\textbf{Estimand Attributes} & \textbf{Description} \\
\hline
\endhead
\hline
\endfoot

\hline
\textbf{Treatment} & We consider a randomized controlled trial with two arms: treatment of interest and placebo. \\
\hline
\textbf{Variable} & Time to each event of interest (both terminal and recurrent) for each subject. \\
\hline
\textbf{Population} & Subjects who satisfy the inclusion and exclusion criteria. \\
\hline
\textbf{Intercurrent Events} & There are two important intercurrent events to consider: treatment discontinuation and death. A treatment policy strategy is applied for treatment discontinuation, meaning that patients are followed for recurrent events regardless of whether they continue the assigned treatment, and their post-discontinuation data are included in the analysis. Death is treated as a terminal event that precludes further recurrent events. \\
\hline
\textbf{Population-Level Summary} & $U(\tau)$ quantifies the mean cumulative burden of recurrent events over the interval $[0, \tau]$ in the study population. The treatment effect can be expressed as either the difference or the ratio of $U(\tau)$ between the two arms. \\
\bottomrule
\end{longtable}

\clearpage

\begin{landscape}
\begin{table}[p]
\centering
\caption{Simulation results for the difference and ratio of AUCs from both unadjusted and covariate-adjusted analyses.}
\scriptsize
\tabcolsep=6pt
\begin{tabular}{ccc ccccc ccccc ccccc ccccc}
\toprule
 &  &  & \multicolumn{10}{c}{Difference of AUC} & \multicolumn{10}{c}{Ratio of AUC} \\ 
\cmidrule(lr){4-13} \cmidrule(lr){14-23}
 &  &  & \multicolumn{5}{c}{Unadjusted} & \multicolumn{5}{c}{Adjusted} & \multicolumn{5}{c}{Unadjusted} & \multicolumn{5}{c}{Adjusted} \\ 
\cmidrule(lr){4-8} \cmidrule(lr){9-13} \cmidrule(lr){14-18} \cmidrule(lr){19-23}
\multicolumn{3}{c}{Settings} &  & \multicolumn{3}{c}{Standard Error (SE)} &  &  & \multicolumn{3}{c}{Standard Error (SE)} &  &  & \multicolumn{3}{c}{Standard Error (SE)} &  &  & \multicolumn{3}{c}{Standard Error (SE)} &  \\ 
\cmidrule(lr){1-3} \cmidrule(lr){5-7} \cmidrule(lr){10-12} \cmidrule(lr){15-17} \cmidrule(lr){20-22}
Cases & Schemes & $\theta$ & Est & Mean & Median & MC & CP & Bias & Mean & Median & MC & CP & Est & Mean & Median & MC & CP & Bias & Mean & Median & MC & CP \\ 
\midrule\addlinespace[4pt]
1 & SPB & -0.32 & -0.229 & 0.049 & 0.049 & 0.047 & 95.8 & 0 & 0.045 & 0.045 & 0.046 & 95.3 & -0.321 & 0.068 & 0.068 & 0.067 & 95.6 & 0 & 0.064 & 0.064 & 0.064 & 94.9 \\ 
1 & SPB & -0.16 & -0.123 & 0.051 & 0.051 & 0.049 & 95.9 & 0 & 0.047 & 0.047 & 0.047 & 95.0 & -0.160 & 0.066 & 0.066 & 0.063 & 95.9 & 0 & 0.061 & 0.061 & 0.061 & 95.1 \\ 
1 & SPB & 0 & -0.001 & 0.053 & 0.053 & 0.051 & 96.0 & 0 & 0.049 & 0.049 & 0.049 & 95.6 & -0.002 & 0.064 & 0.064 & 0.061 & 96.0 & 0 & 0.059 & 0.059 & 0.058 & 95.5 \\ [1ex]
1 & Simple & -0.32 & -0.228 & 0.049 & 0.049 & 0.048 & 95.4 & 0 & 0.045 & 0.045 & 0.045 & 95.0 & -0.320 & 0.068 & 0.068 & 0.068 & 95.2 & 0 & 0.064 & 0.064 & 0.064 & 94.7 \\ 
1 & Simple & -0.16 & -0.123 & 0.051 & 0.051 & 0.051 & 95.3 & 0 & 0.047 & 0.047 & 0.046 & 95.1 & -0.160 & 0.066 & 0.066 & 0.066 & 95.3 & 0 & 0.061 & 0.061 & 0.060 & 95.1 \\ 
1 & Simple & 0 & 0 & 0.053 & 0.053 & 0.053 & 95.4 & 0 & 0.049 & 0.049 & 0.049 & 95.0 & 0 & 0.064 & 0.064 & 0.064 & 95.4 & 0 & 0.059 & 0.059 & 0.059 & 95.0 \\ [1.5ex]
2 & SPB & -0.32 & -0.196 & 0.050 & 0.049 & 0.048 & 95.6 & 0 & 0.046 & 0.046 & 0.046 & 95.1 & -0.267 & 0.068 & 0.068 & 0.066 & 95.7 & 0 & 0.063 & 0.063 & 0.063 & 95.0 \\ 
2 & SPB & -0.16 & -0.104 & 0.051 & 0.051 & 0.049 & 96.0 & 0 & 0.047 & 0.047 & 0.047 & 94.9 & -0.134 & 0.066 & 0.066 & 0.063 & 96.0 & 0 & 0.061 & 0.061 & 0.060 & 95.0 \\ 
2 & SPB & 0 & -0.001 & 0.053 & 0.053 & 0.051 & 96.0 & 0 & 0.049 & 0.049 & 0.049 & 95.6 & -0.002 & 0.064 & 0.064 & 0.061 & 96.0 & 0 & 0.059 & 0.059 & 0.058 & 95.5 \\ [1ex]
2 & Simple & -0.32 & -0.195 & 0.050 & 0.050 & 0.049 & 95.5 & 0 & 0.046 & 0.046 & 0.046 & 95.0 & -0.267 & 0.068 & 0.068 & 0.067 & 95.2 & 0 & 0.063 & 0.063 & 0.063 & 95.0 \\ 
2 & Simple & -0.16 & -0.104 & 0.051 & 0.051 & 0.051 & 95.3 & 0 & 0.047 & 0.047 & 0.047 & 95.2 & -0.133 & 0.066 & 0.066 & 0.065 & 95.3 & 0 & 0.061 & 0.061 & 0.060 & 95.3 \\ 
2 & Simple & 0 & 0 & 0.053 & 0.053 & 0.053 & 95.4 & 0 & 0.049 & 0.049 & 0.049 & 95.0 & 0 & 0.064 & 0.064 & 0.064 & 95.4 & 0 & 0.059 & 0.059 & 0.059 & 95.0 \\ [1.5ex]
3 & SPB & -0.32 & -0.168 & 0.050 & 0.050 & 0.048 & 96.1 & 0 & 0.046 & 0.046 & 0.046 & 95.3 & -0.225 & 0.067 & 0.067 & 0.064 & 95.8 & 0 & 0.062 & 0.062 & 0.061 & 95.4 \\ 
3 & SPB & -0.16 & -0.090 & 0.052 & 0.052 & 0.049 & 95.9 & 0 & 0.048 & 0.048 & 0.047 & 95.5 & -0.114 & 0.065 & 0.065 & 0.063 & 95.7 & 0 & 0.060 & 0.060 & 0.060 & 95.5 \\ 
3 & SPB & 0 & -0.001 & 0.053 & 0.053 & 0.051 & 96.0 & 0 & 0.049 & 0.049 & 0.049 & 95.6 & -0.002 & 0.064 & 0.064 & 0.061 & 96.0 & 0 & 0.059 & 0.059 & 0.058 & 95.5 \\ [1ex]
3 & Simple & -0.32 & -0.167 & 0.050 & 0.050 & 0.050 & 95.3 & 0 & 0.046 & 0.046 & 0.046 & 95.2 & -0.223 & 0.067 & 0.067 & 0.066 & 95.5 & 0 & 0.062 & 0.062 & 0.062 & 94.9 \\ 
3 & Simple & -0.16 & -0.089 & 0.052 & 0.052 & 0.051 & 95.1 & 0 & 0.048 & 0.048 & 0.047 & 95.1 & -0.113 & 0.065 & 0.065 & 0.065 & 95.2 & 0 & 0.060 & 0.060 & 0.060 & 95.0 \\ 
3 & Simple & 0 & 0 & 0.053 & 0.053 & 0.053 & 95.4 & 0 & 0.049 & 0.049 & 0.049 & 95.0 & 0 & 0.064 & 0.064 & 0.064 & 95.4 & 0 & 0.059 & 0.059 & 0.059 & 95.0 \\ [1.5ex]
4 & SPB & -0.32 & -0.126 & 0.051 & 0.051 & 0.049 & 95.8 & 0 & 0.047 & 0.047 & 0.046 & 95.3 & -0.163 & 0.066 & 0.066 & 0.063 & 95.8 & 0 & 0.061 & 0.061 & 0.060 & 95.3 \\ 
4 & SPB & -0.16 & -0.067 & 0.052 & 0.052 & 0.050 & 96.0 & 0 & 0.048 & 0.048 & 0.047 & 95.4 & -0.084 & 0.065 & 0.065 & 0.062 & 96.0 & 0 & 0.060 & 0.060 & 0.059 & 95.5 \\ 
4 & SPB & 0 & -0.001 & 0.053 & 0.053 & 0.051 & 96.0 & 0 & 0.049 & 0.049 & 0.049 & 95.6 & -0.002 & 0.064 & 0.064 & 0.061 & 96.0 & 0 & 0.059 & 0.059 & 0.058 & 95.5 \\ [1ex]
4 & Simple & -0.32 & -0.125 & 0.051 & 0.051 & 0.050 & 95.1 & 0 & 0.047 & 0.047 & 0.047 & 95.0 & -0.162 & 0.066 & 0.066 & 0.065 & 95.1 & 0 & 0.061 & 0.061 & 0.061 & 94.8 \\ 
4 & Simple & -0.16 & -0.065 & 0.052 & 0.052 & 0.052 & 95.3 & 0 & 0.048 & 0.048 & 0.048 & 95.1 & -0.082 & 0.065 & 0.065 & 0.064 & 95.4 & 0 & 0.060 & 0.060 & 0.060 & 95.0 \\ 
4 & Simple & 0 & 0 & 0.053 & 0.053 & 0.053 & 95.4 & 0 & 0.049 & 0.049 & 0.049 & 95.0 & 0 & 0.064 & 0.064 & 0.064 & 95.4 & 0 & 0.059 & 0.059 & 0.059 & 95.0 \\ [1.5ex]
5 & SPB & -0.32 & -0.245 & 0.036 & 0.036 & 0.033 & 96.9 & 0 & 0.030 & 0.030 & 0.030 & 95.0 & -0.299 & 0.044 & 0.044 & 0.041 & 96.8 & 0 & 0.037 & 0.037 & 0.037 & 95.1 \\ 
5 & SPB & -0.16 & -0.123 & 0.037 & 0.037 & 0.034 & 96.5 & 0 & 0.031 & 0.031 & 0.031 & 95.2 & -0.139 & 0.042 & 0.042 & 0.038 & 96.3 & 0 & 0.035 & 0.035 & 0.035 & 95.2 \\ 
5 & SPB & 0 & 0 & 0.038 & 0.038 & 0.036 & 96.5 & 0 & 0.032 & 0.032 & 0.033 & 94.8 & 0 & 0.040 & 0.040 & 0.038 & 96.6 & 0 & 0.034 & 0.034 & 0.035 & 94.9 \\ [1ex]
5 & Simple & -0.32 & -0.244 & 0.036 & 0.036 & 0.036 & 95.0 & 0 & 0.030 & 0.030 & 0.030 & 95.1 & -0.298 & 0.044 & 0.044 & 0.044 & 94.9 & 0 & 0.037 & 0.037 & 0.037 & 95.0 \\ 
5 & Simple & -0.16 & -0.122 & 0.037 & 0.037 & 0.038 & 94.5 & 0 & 0.031 & 0.031 & 0.032 & 94.2 & -0.138 & 0.042 & 0.042 & 0.043 & 94.5 & 0 & 0.035 & 0.035 & 0.036 & 94.2 \\ 
5 & Simple & 0 & 0 & 0.038 & 0.038 & 0.038 & 94.4 & 0 & 0.032 & 0.032 & 0.033 & 94.9 & 0 & 0.040 & 0.040 & 0.040 & 94.4 & 0 & 0.034 & 0.034 & 0.034 & 94.9 \\ 
\bottomrule
\end{tabular}
\begin{tablenotes}
\scriptsize
\item[] \textbf{Notes:}
\begin{itemize}
    \item ``Est'' is the Monte Carlo mean of the estimated unadjusted differences and ratios of AUCs.
    \item ``Bias'' is the difference between the Monte Carlo mean of the adjusted estimates and the corresponding ``Est.''
    \item ``Mean'' and ``Median'' denote the mean and median of the estimated standard errors for each point estimate---$\Delta \hat U(\tau)$, $\Delta \hat U^{(a)}(\tau)$, $\hat R(\tau)$, and $\hat R^{(a)}(\tau)$.
    \item ``MC'' represents the Monte Carlo standard deviation of the corresponding point estimates.
    \item Coverage probability (CP) refers to the likelihood that the 95\% confidence interval (CI) contains the respective ``Est.''
\end{itemize}
\end{tablenotes}
\label{tab:auc_results}
\end{table}
\end{landscape}

\clearpage

\begin{landscape}
\begin{table}[p]
\centering
\caption{Estimation results for the ratio and the difference of AUCs between the two arms. The unadjusted estimates are $\hat R(\tau)= \hbox{log} \frac{\hat U_1(\tau)}{\hat U_0(\tau)}$ and $\Delta \hat U(\tau) = \hat U_1(\tau) - \hat U_0(\tau)$, where $\hat U_1(\tau)$ and $\hat U_0(\tau)$ are the AUC estimates for the treatment arm and the placebo arm, respectively. The adjusted estimates are $\hat R^{(a)}(\tau)=\hat R(\tau) - \frac{n}{n_0 n_1} \hat{\mathcal{A}}_{R}$ and $\Delta \hat U^{(a)}(\tau)=\Delta \hat U(\tau) - \frac{n}{\sqrt{n_0}\sqrt{n_1}} \hat{\mathcal{A}}_{\Delta}$, with $\hat{\mathcal{A}}_{R}$ and $\hat{\mathcal{A}}_{\Delta}$ representing the adjustment. \label{hfactiontable}}
{
\begin{tabular}{lccccccccc}
\toprule
     & \multicolumn{1}{c}{} & \multicolumn{4}{c}{Unadjusted}        & \multicolumn{4}{c}{Adjusted}          \\ \cmidrule(lr){3-6}  \cmidrule(lr){7-10} \addlinespace[4pt]
\multicolumn{10}{c}{\textbf{\textit{Ratio of AUCs}}} \\ [2.5ex]
Studies & $\tau$                  & $\hbox{exp}\{\hat R(\tau)\}$ & $\hbox{Var}\{\hat R(\tau)\}$    & 95\% CI        & p    & $\hbox{exp}\{\hat R^{(a)}(\tau)\}$ & $\hbox{Var}\{\hat R^{(a)}(\tau)\}$     & 95\% CI        & p    \\ \hline \\
HF-ACTION  & 4                    & 0.886 & 0.0151 & (0.696, 1.127) & 0.32 & 0.862 & 0.0147 & (0.679, 1.093) & 0.22 \\ \\
COPERNICUS  & 2.2                    & 0.818 & 0.0055 & (0.707, 0.946) & 0.007 & 0.805 & 0.0052 & (0.698, 0.928) & 0.003 \\ \\
\multicolumn{10}{c}{\textbf{\textit{Difference of AUCs}}} \\[2.5ex]
Studies & $\tau$                  & $\Delta \hat U(\tau)$ & $\hbox{Var}\{\Delta \hat U(\tau)\}$    & 95\% CI        & p    & $\Delta \hat U^{(a)}(\tau)$ & $\hbox{Var}\{\Delta \hat U^{(a)}(\tau)\}$     & 95\% CI        & p    \\ \hline \\
HF-ACTION  & 4                    & -0.874 & 0.7695 & (-2.594, 0.845) & 0.32 & -1.071 & 0.7526 & (-2.772, 0.629) & 0.22 \\ \\
COPERNICUS  & 2.2                    & -0.335 & 0.0153 & (-0.578, -0.093) & 0.007 & -0.363 & 0.0146 & (-0.600, -0.126) & 0.003 \\
\bottomrule
\end{tabular}}
\end{table}
\end{landscape}

\end{document}


\maketitle

\section{Covariate Adjustment for the Difference of AUCs}\label{app:sec:cov_diff}

Let $\Delta \hat U(\tau) = \hat U_1(\tau) - \hat U_0(\tau)$ and $\hat{\mathcal{U}}_{L, \Delta}=\frac{\sqrt{n_0}\sqrt{n_1}}{n} \left\{ \Delta \hat U(\tau) - \Delta U(\tau) \right\} $, where $n_0$ and $n_1$ represent the sample sizes for the placebo and treatment arms, respectively. The linearization of $\hat{\mathcal{U}}_{L,\Delta}$ is as follows:
\begin{eqnarray*}
    \hat{\mathcal{U}}_{L,\Delta} &=& \frac{\sqrt{n_0}\sqrt{n_1}}{n} \left\{ \hat U_{1}(\tau) -  U_{1}(\tau) \right\} - \frac{\sqrt{n_0}\sqrt{n_1}}{n} \left\{  \hat U_{0}(\tau) - U_{0}(\tau) \right\} \\
    &=& \frac{1}{n} \left( \frac{n_0}{n_1} \right)^{\frac{1}{2}} \sum_{i=1}^{n_1} \hat \psi_{i}(\tau) - \frac{1}{n} \left( \frac{n_1}{n_0} \right)^{\frac{1}{2}} \sum_{j=1}^{n_0} \hat \psi_{j}(\tau) +o_p(1) \\
    &=& \frac{1}{n} \sum_{i=1}^{n} \left\{ I_{i}  \left( \frac{n_0}{n_1} \right)^{\frac{1}{2}}   \hat \psi_{i}(\tau) - (1-I_i)  \left( \frac{n_1}{n_0} \right)^{\frac{1}{2}}   \hat \psi_{i}(\tau) \right\} +o_p(1)\\
    &=& \frac{1}{n} \sum_{i=1}^{n} \left\{ I_{i}  (\hat P^{\Delta}_{i} - \hat Q^{\Delta}_{i}) - (1-I_i)  (\hat P^{\Delta}_{i} - \hat Q^{\Delta}_{i}) \right\} +o_p(1)
\end{eqnarray*}
where $I_{i}=1$ if subject $i$ is in the treatment arm and 0 otherwise, $\hat P^{\Delta}_{i} = \left( \frac{n_1}{n_0} \right)^{\frac{1}{2}  (-1)^{I_i}} \hat P_i$ and $\hat Q^{\Delta}_{i} = \left( \frac{n_1}{n_0} \right)^{\frac{1}{2}  (-1)^{I_i}} \hat Q_i$.

The unadjusted test statistic on the difference of AUCs is $\mathcal{T}_{L,\Delta} = \sqrt{n}  \hat{\mathcal{U}}_{L, \Delta}/\hat{\sigma}_{L, \Delta}$, where 
$$\hat{\sigma}_{L, \Delta}^{2} = \frac{1}{n}  \left[ n_0  \mathbb{E}\left\{ \hat \psi_{1}(\tau)^{T} \hat \psi_{1}(\tau) \right\} + n_1  \mathbb{E}\left\{ \hat \psi_{0}(\tau)^{T} \hat \psi_{0}(\tau)\right\} \right] .$$
Here, $\hat \psi_{1}(\tau)$ and $\hat \psi_{0}(\tau)$ consist of $\hat \psi_{i}(\tau)$ from subjects in the treatment and placebo arms, respectively.

\begin{theorem}
    The following results hold only under the simple randomization scheme: 

    Under the null hypothesis \( H_{0, \Delta}: \Delta U(\tau) = 0 \) or the alternative hypothesis, we have
    \[
        \sqrt{n} \hat{\mathcal{U}}_{L,\Delta} \stackrel{d}{\to} N\left(0, \sigma_{L, \Delta}^{2}\right), \quad
        \hat{\sigma}_{L, \Delta}^{2} \stackrel{p}{\to} \sigma_{L, \Delta}^{2}, \quad \text{and} \quad
        \mathcal{T}_{L, \Delta} \stackrel{d}{\to} N(0, 1),
    \]
    where 
    \[
        \sigma_{L, \Delta}^{2} = \frac{1}{n} \left[ n_0 \mathbb{E}\left\{ \psi_{1}(\tau)^{T} \psi_{1}(\tau) \right\} + n_1 \mathbb{E}\left\{ \psi_{0}(\tau)^{T} \psi_{0}(\tau) \right\} \right].
    \]
    \label{appenthm_diff}
\end{theorem}

\begin{proof}
By the Central Limit Theorem, we have $\sqrt{n}  \hat{\mathcal{U}}_{L,\Delta}\stackrel{d}{\to} N\left(0, \sigma_{L, \Delta}^{2}\right) $. Additionally, by the Law of Large Numbers, $\hat{\sigma}_{L, \Delta}^{2} \stackrel{p}{\to} \sigma_{L, \Delta}^{2}$. Applying Slutsky’s Theorem, we then have, $\mathcal{T}_{L, \Delta} \stackrel{d}{\to} N(0, 1) $.
\end{proof}

Thus, $H_{0, \Delta}$ is rejected when $|\mathcal{T}_{L,\Delta}| > z_{\alpha/2}$, where $\alpha$ is a given significance level and $z_{\alpha/2}$ is the $(1-\alpha/2)$th quantile of the standard normal distribution. The unadjusted $100(1-\alpha) \%$ confidence interval (CI) for $\Delta U(\tau)$ is $\big( \Delta \hat U(\tau) - z_{\alpha/2}  SE\big\{\Delta \hat U(\tau)\big\}, \Delta \hat U(\tau) + z_{\alpha/2}  SE\big\{\Delta \hat U(\tau)\big\} \big)$, where $SE\left\{ \Delta \hat U(\tau) \right\} = \left[ \frac{1}{n_1}  \mathbb{E}\left\{ \hat \psi_{1}(\tau)^{T} \hat \psi_{1}(\tau) \right\} + \frac{1}{n_0}  \mathbb{E}\left\{ \hat \psi_{0}(\tau)^{T} \hat \psi_{0}(\tau) \right\}  \right]^{1/2}$.

Let $\boldsymbol{X}^{*}$ be a $p$-dimensional vector of covariates. Define $X_{i}=X_{i}^{*}-\bar X$, where $X_{i}^{*}$ is the $i$th column of $\boldsymbol{X}^{*}$, and $\bar{X}$ is the sample mean of all the $X_{i}^{*}$, ensuring $\mathbb{E}(X_i)=0$. Following \cite{ye2024covariate}, we treat $\hat P^{\Delta}_{i}$ and $\hat Q^{\Delta}_{i}$ as outcomes and apply the generalized regression adjustment or augmentation \citep{tsiatis2008covariate}. This results in the covariate-adjusted $\hat{\mathcal{U}}_{CL,\Delta}$:
\begin{eqnarray*}
    \hat{\mathcal{U}}_{CL,\Delta} = \frac{1}{n} \sum_{i=1}^{n} \Bigg[ I_{i} \left\{ (\hat P^{\Delta}_{i} - X_{i}^{T} \hat \beta_{1,P}^{\Delta} ) 
    - (\hat Q^{\Delta}_{i} - X_{i}^{T} \hat \beta_{1,Q}^{\Delta}) \right\}  - (1 - I_i) \left\{ (\hat P^{\Delta}_{i} - X_{i}^{T} \hat \beta_{0,P}^{\Delta}) - (\hat Q^{\Delta}_{i} - X_{i}^{T} \hat \beta_{0,Q}^{\Delta}) \right\} \Bigg] = \hat{\mathcal{U}}_{L,\Delta} - \hat{\mathcal{A}}_{\Delta}
\end{eqnarray*}
where
\begin{eqnarray*}
    \hat{\mathcal{A}}_{\Delta} &=& \frac{1}{n} \sum_{i=1}^{n} \Bigg\{ I_i  X_{i}^{T}  \left( \hat \beta_{1,P}^{\Delta} - \hat \beta_{1,Q}^{\Delta} \right) 
    - (1-I_i) X_{i}^{T} \left( \hat \beta_{0,P}^{\Delta} - \hat \beta_{0,Q}^{\Delta} \right) \Bigg\}, \\
    \hat \beta_{j,P}^{\Delta} &=& \left( \sum_{i:I_i=j} X_i X_{i}^{T} \right)^{-1} 
    \sum_{i:I_i=j} X_{i} \hat P^{\Delta}_{i}, \\
    \hat \beta_{j,Q}^{\Delta} &=& \left( \sum_{i:I_i=j} X_i X_{i}^{T} \right)^{-1} 
    \sum_{i:I_i=j} X_{i} \hat Q^{\Delta}_{i}.
\end{eqnarray*}

Denote the adjusted $\Delta \hat U(\tau)$ by $\Delta \hat U^{(a)}(\tau)=\Delta \hat U(\tau) - \frac{n}{\sqrt{n_0}\sqrt{n_1}} \hat{\mathcal{A}}_{\Delta}$. Then $ \hat{\mathcal{U}}_{CL, \Delta}=\frac{\sqrt{n_0}\sqrt{n_1}}{n} \{ \Delta \hat U^{(a)}(\tau) - \Delta U(\tau) \} $. Define the covariate-adjusted test statistic on the difference of AUCs as $\mathcal{T}_{CL,\Delta} = \sqrt{n}  \hat{\mathcal{U}}_{CL, \Delta}/\hat{\sigma}_{CL, \Delta}$, where $\hat{\sigma}_{CL, \Delta}^2= \hat{\sigma}_{L, \Delta}^2 - \frac{n_0 n_1}{n^2} (\hat \beta_{1,P}^{\Delta} - \hat \beta_{1,Q}^{\Delta} + \hat \beta_{0,P}^{\Delta} - \hat \beta_{0,Q}^{\Delta})^{T}   \sum_{X}  (\hat \beta_{1,P}^{\Delta} - \hat \beta_{1,Q}^{\Delta} + \hat \beta_{0,P}^{\Delta} - \hat \beta_{0,Q}^{\Delta}) $. The theorem below establishes the asymptotic properties of $\hat{\mathcal{U}}_{CL,\Delta}$ and $\mathcal{T}_{CL, \Delta}$. The technical proofs follow analogously from the proof presented in Section \ref{sec:thm1proof} of the Supplementary Materials.
\begin{theorem}\label{theoadjdif}
    The following results hold under both simple randomization and covariate-adaptive randomization schemes:
    \\
    Under the $H_{0, \Delta}: \Delta U(\tau)=0$ or the alternative hypothesis, $\sqrt{n}  \hat{\mathcal{U}}_{CL,\Delta}\stackrel{d}{\to} N\left(0, \sigma_{CL, \Delta}^{2}\right) $, $\hat \sigma_{CL, \Delta}^{2} \stackrel{p}{\to} \sigma_{CL, \Delta}^{2}$, and $\mathcal{T}_{CL, \Delta} \stackrel{d}{\to} N(0, 1) $, where $\sigma_{CL, \Delta}^2= \sigma_{L, \Delta}^2 - \frac{n_0 n_1}{n^2} ( \beta_{1,P}^{\Delta} - \beta_{1,Q}^{\Delta} + \beta_{0,P}^{\Delta} - \beta_{0,Q}^{\Delta})^{T}   \sum_{X}  ( \beta_{1,P}^{\Delta} - \beta_{1,Q}^{\Delta} + \beta_{0,P}^{\Delta} - \beta_{0,Q}^{\Delta}) $. 
\end{theorem}

$H_{0,\Delta}$ is rejected when $|\mathcal{T}_{CL,\Delta}| > z_{\alpha/2}$ and the adjusted $100 (1-\alpha) \%$ CI for $\Delta  U(\tau)$ is $( \Delta \hat U^{(a)}(\tau) - z_{\alpha/2}  SE\left\{\Delta \hat U^{(a)}(\tau)\right\}, \Delta \hat U^{(a)}(\tau) + z_{\alpha/2}  SE\left\{\Delta \hat U^{(a)}(\tau)\right\} )$, where $SE\left\{ \Delta \hat U^{(a)}(\tau)\right\} = \left( \frac{n}{n_0 n_1}  \hat{\sigma}_{CL, \Delta}^2  \right)^{1/2}$.

As proved by \cite{ye2024covariate}, $\sigma_{CL, \Delta}^2 \leq \sigma_{L, \Delta}^2$, with strict inequality holding unless either (i) the covariates $X_{i}^{*}$ are uncorrelated with $P_{i}^{\Delta}$ and $Q_{i}^{\Delta}$, or (ii) the covariates not used in the randomization are uncorrelated with $P_{i}^{\Delta}$ and $Q_{i}^{\Delta}$, given the covariates used in the randomization. Therefore, the adjusted $\mathcal{T}_{CL, \Delta}$ provides a guaranteed efficiency gain over the unadjusted $\mathcal{T}_{L, \Delta}$.

\section{Linearization and Asymptotic Properties of Unadjusted Test Statistics for the Ratio of AUCs}\label{sec:ratioauc}

Let $\hat R(\tau)= \hbox{log} \frac{\hat U_1(\tau)}{\hat U_0(\tau)}$ and $\hat{\mathcal{U}}_{L, R} = \frac{n_0 n_1}{n} \left\{ \hat R(\tau) - R(\tau) \right\} $. The linearization of $\hat{\mathcal{U}}_{L, R}$ is as follows:
\begin{eqnarray*}
    \hat{\mathcal{U}}_{L, R} &=& \frac{n_0  n_1}{n} \left\{ \hbox{log} \hat U_{1}(\tau) - \hbox{log} U_{1}(\tau) \right\} - \frac{n_0  n_1}{n} \left\{ \hbox{log} \hat U_{0}(\tau) - \hbox{log} U_{0}(\tau) \right\} \\
    &=& \frac{n_0 n_1}{n}  \frac{1}{\hat U_{1}(\tau)}  \left\{ \hat U_{1}(\tau) -  U_{1}(\tau) \right\} - \frac{n_0 n_1}{n}  \frac{1}{\hat U_{0}(\tau)}  \left\{ \hat U_{0}(\tau) -  U_{0}(\tau) \right\} +o_p(1)\\
    &=&  \frac{n_0 n_1}{n}  \frac{1}{\hat U_{1}(\tau)}  \frac{1}{n_1} \sum_{i=1}^{n_1} \hat \psi_{i}(\tau) - \frac{n_0 n_1}{n}  \frac{1}{\hat U_{0}(\tau)}  \frac{1}{n_0} \sum_{j=1}^{n_0} \hat \psi_{j}(\tau) +o_p(1)\\
    & = & \frac{1}{n} \sum_{i=1}^{n} \left\{ I_i  \frac{n_0}{\hat U_{1}(\tau)}  \psi_{i}(\tau) - (1-I_i)  \frac{n_1}{\hat U_{0}(\tau)}  \psi_{i}(\tau)  \right\} +o_p(1)\\
    &=& \frac{1}{n} \sum_{i=1}^{n} \left\{ I_{i}  (\hat P^{R}_{i} - \hat Q^{R}_{i}) - (1-I_i)  (\hat P^{R}_{i} - \hat Q^{R}_{i}) \right\} +o_p(1)
\end{eqnarray*}
where $\hat P^{R}_{i} = \left\{ \frac{n_0}{\hat U_{1}(\tau)}  \right\}^{I_i}  \left\{ \frac{n_1}{\hat U_{0}(\tau)} \right\}^{1-I_i}   \hat P_i$ and $\hat Q^{R}_{i} = \left\{ \frac{n_0}{\hat U_{1}(\tau)}  \right\}^{I_i}  \left\{ \frac{n_1}{\hat U_{0}(\tau)} \right\}^{1-I_i}   \hat Q_i$.

Define the unadjusted test statistic on the ratio of AUCs as $\mathcal{T}_{L,R} = \sqrt{n}  \hat{\mathcal{U}}_{L, R}/\hat{\sigma}_{L, R}$, where 
$$\hat{\sigma}_{L, R}^{2} = \frac{1}{n}  \left[ \frac{n_0^2 n_1}{\hat U_{1}(\tau)^{2}}  \mathbb{E}\left\{ \hat \psi_{1}(\tau)^{T} \hat \psi_{1}(\tau) \right\} + \frac{n_1^2 n_0}{\hat U_{0}(\tau)^{2}}  \mathbb{E}\left\{ \hat \psi_{0}(\tau)^{T} \hat \psi_{0}(\tau)\right\} \right] $$

\begin{theorem}
    The following results hold only under the simple randomization scheme: 

    Under the null hypothesis \( H_{0,R}: R(\tau) = 0 \) or the alternative hypothesis, we have
    \[
        \sqrt{n} \hat{\mathcal{U}}_{L,R} \stackrel{d}{\to} N\left(0, \sigma_{L, R}^{2}\right), \quad
        \hat{\sigma}_{L, R}^{2} \stackrel{p}{\to} \sigma_{L, R}^{2}, \quad \text{and} \quad
        \mathcal{T}_{L, R} \stackrel{d}{\to} N(0, 1),
    \]
    where 
    \[
        \sigma_{L, R}^{2} = \frac{1}{n} \left[ \frac{n_0^2 n_1}{U_{1}(\tau)^{2}} \mathbb{E}\left\{ \psi_{1}(\tau)^{T} \psi_{1}(\tau) \right\} + \frac{n_1^2 n_0}{U_{0}(\tau)^{2}} \mathbb{E}\left\{ \psi_{0}(\tau)^{T} \psi_{0}(\tau) \right\} \right].
    \]
\end{theorem}

The theorem can be proved similarly to the proof of Theorem \ref{appenthm_diff}. 

Thus, $H_{0,R}$ is rejected when $|\mathcal{T}_{L,R}| > z_{\alpha/2}$. The unadjusted $100(1-\alpha) \%$ CI for $\hat R(\tau)$ is $\big(\frac{\hat U_1(\tau)}{\hat U_0(\tau)} \hbox{exp}\big[ - z_{\alpha/2}  SE\big\{ \hat R(\tau) \big\} \big], \frac{\hat U_1(\tau)}{\hat U_0(\tau)} \hbox{exp}\big[ z_{\alpha/2}  SE\big\{ \hat R(\tau) \big\} \big]\big)$, where $SE\left\{ \hat R(\tau) \right\} = \left[ \frac{1}{n_1 \hat U_{1}(\tau)^{2}}  \mathbb{E}\left\{ \hat \psi_{1}(\tau)^{T} \hat \psi_{1}(\tau) \right\} +  \frac{1}{n_0 \hat U_{0}(\tau)^{2}}  \mathbb{E}\left\{ \hat \psi_{0}(\tau)^{T} \hat \psi_{0}(\tau) \right\} \right]^{1/2}$.

\section{Proof of Theorem 1 in the Main Text}\label{sec:thm1proof}

The proof is similar to that of Theorem 1 in \cite{ye2024covariate}. Below is a brief outline of how $\hat \sigma_{CL, R}^2$ is derived.
\begin{eqnarray*}
    \hat{\mathcal{U}}_{CL,R} &=& \frac{1}{n} \sum_{i=1}^{n} \Bigl( I_{i}  \bigl[ \{\hat P^{R}_{i} - (X_{i}^{*}-\bar X)^{T} \hat \beta_{1,P}^{R} \} - \{\hat Q^{R}_{i} - (X_{i}^{*}-\bar X)^{T} \hat \beta_{1,Q}^{R}\} \bigr] \\
    &-& (1-I_i)  \bigl[ \{\hat P^{R}_{i} - (X_{i}^{*}-\bar X)^{T} \hat \beta_{0,P}^{R}\} - \{ \hat Q^{R}_{i} - (X_{i}^{*}-\bar X)^{T} \hat \beta_{0,Q}^{R}\} \bigr] \Bigr)  \\
    &=& \frac{1}{n} \sum_{i=1}^{n} \Bigl\{ I_{i}  \Bigl( \bigl[\hat P^{R}_{i} - \{X_{i}^{*} -\mu_{X} - (\bar X -\mu_{X} )\}^{T} \hat \beta_{1,P}^{R} \bigr] - \bigl[\hat Q^{R}_{i} - \{X_{i}^{*} -\mu_{X} - (\bar X -\mu_{X} )\}^{T} \hat \beta_{1,Q}^{R}\bigr] \Bigr) \\
    && - (1-I_i)  \Bigl( \bigl[\hat P^{R}_{i} - \{X_{i}^{*} -\mu_{X} - (\bar X -\mu_{X} )\}^{T} \hat \beta_{0,P}^{R}\bigr] - \bigl[ \hat Q^{R}_{i} - \{X_{i}^{*} -\mu_{X} - (\bar X -\mu_{X} )\}^{T} \hat \beta_{0,Q}^{R}\bigr] \Bigr) \Bigr\} \\
    &=& M_1 - M_2 + M_3.
\end{eqnarray*}
where 
\begin{eqnarray*}
    M_1&=&\frac{1}{n} \sum_{i=1}^{n} (1-I_i)  \bigl[ \{\hat P^{R}_{i} - (X_{i}^{*}-\mu_X)^{T} \hat \beta_{0,P}^{R}\} - \{ \hat Q^{R}_{i} - (X_{i}^{*}-\mu_X)^{T} \hat \beta_{0,Q}^{R}\} \bigr] \\
    M_2&=&\frac{1}{n} \sum_{i=1}^{n} I_{i}  \bigl[ \{\hat P^{R}_{i} - (X_{i}^{*}-\bar X)^{T} \hat \beta_{1,P}^{R} \} - \{\hat Q^{R}_{i} - (X_{i}^{*}-\bar X)^{T} \hat \beta_{1,Q}^{R}\} \bigr] \\
    M_3&=&(\bar X -\mu_{X})  \left\{ \pi  (\hat \beta_{0,P}^{R} - \hat \beta_{0,Q}^{R}) -(1-\pi)  (\hat \beta_{1,P}^{R} - \hat \beta_{1,Q}^{R}) \right\}
\end{eqnarray*}
with $\pi=\frac{n_0}{n}$.

By using the definition $\beta_{0,P}^{R}=\sum_{X}^{-1}cov(X_i^*, P_i^R)$, we have 
$$\mathbb{E}\left[ X_i^{*T} \left\{ P^{R}_{i} - (X_{i}^{*}-\mu_X)^{T} \beta_{0,P}^{R} \right\} \right]=cov(X_i^{*}, P^{R}_{i})-cov(X_i^*, P_i^R)=0$$
Similarly, 
$\mathbb{E}\left[ X_i^{*T} \left\{ Q^{R}_{i} - (X_{i}^{*}-\mu_X)^{T} \beta_{0,Q}^{R} \right\} \right]=0$. Therefore, 
$cov(M_1, M_3) \propto cov(M_1, \bar X)=\frac{1}{n} \sum_{i=1}^{n} cov\left(M_1, X_i \right)=0$. Similarly, $cov(M_1, M_3)=0$. Then, 
\begin{eqnarray*}
   \hat \sigma_{CL, R}^2 - \hat \sigma_{L, R}^2 &=& -\hbox{Var}\left\{ (\bar X - \mu_X)^{T} \hat \beta_{0}  \right\} - \hbox{Var}\left\{ (\bar X - \mu_X)^{T} \hat \beta_{1}  \right\} \\
   && + \left\{ \pi \hat \beta_{0}^{R} - (1-\pi) \hat \beta_{1}^{R}  \right\}^{T} \hbox{Var}(\bar X - \mu_X) \left\{ \pi \hat \beta_{0}^{R} - (1-\pi) \hat \beta_{1}^{R}  \right\} \\
   &=& -\pi (1-\pi) (\hat \beta_{0}^{R}+ \hat \beta_{1}^{R})^{T} \sum_X (\hat \beta_{0}^{R}+ \hat \beta_{1}^{R}).
\end{eqnarray*}
where $\hat \beta_{0}^{R}=\hat \beta_{0,P}^{R} - \hat \beta_{0,Q}^{R}$ and $\hat \beta_{1}^{R}=\hat \beta_{1,P}^{R} - \hat \beta_{1,Q}^{R}$.

\section{Additional simulation results}\label{sec:simresults}
\subsection{Simulation Results for a Small Sample Size (Total of 400 Subjects)\label{sec:results_small}}

Table \ref{tab:auc_results_small} presents simulation results for the difference and ratio of AUCs from both unadjusted and covariate-adjusted analyses, across various scenarios and randomization schemes, based on 5,000 simulations with a total sample size of 400 subjects. The ``SE'' of ``Est'' and ``Bias'' is calculated as the corresponding ``MC'' divided by $\sqrt{5000}$. The results are comparable to those in Table 2.

\begin{landscape}
\begin{table}[htbp]
\centering
\caption{Simulation results for the difference and ratio of AUCs from unadjusted and covariate-adjusted analyses across various scenarios and randomization schemes, based on a total sample size of 400 subjects.}
\scriptsize
\tabcolsep=4pt
\begin{tabular}{ccc ccccc ccccc ccccc ccccc}
\toprule
 &  &  & \multicolumn{10}{c}{Difference of AUC} & \multicolumn{10}{c}{Ratio of AUC} \\ 
\cmidrule(lr){4-13} \cmidrule(lr){14-23}
 &  &  & \multicolumn{5}{c}{Unadjusted} & \multicolumn{5}{c}{Adjusted} & \multicolumn{5}{c}{Unadjusted} & \multicolumn{5}{c}{Adjusted} \\ 
\cmidrule(lr){4-8} \cmidrule(lr){9-13} \cmidrule(lr){14-18} \cmidrule(lr){19-23}
\multicolumn{3}{c}{Settings} &  & \multicolumn{3}{c}{Standard Error (SE)} &  &  & \multicolumn{3}{c}{Standard Error (SE)} &  &  & \multicolumn{3}{c}{Standard Error (SE)} &  &  & \multicolumn{3}{c}{Standard Error (SE)} &  \\ 
\cmidrule(lr){1-3} \cmidrule(lr){5-7} \cmidrule(lr){10-12} \cmidrule(lr){15-17} \cmidrule(lr){20-22}
Cases & Schemes & $\theta$ & Est (SE) & Mean & Median & MC & CP & Bias (SE) & Mean & Median & MC & CP & Est (SE) & Mean & Median & MC & CP & Bias (SE) & Mean & Median & MC & CP \\ 
\midrule\addlinespace[4pt]
 1 & SPB & -0.32 & -0.228 (0.002) & 0.109 & 0.109 & 0.107 & 95.2 & 0 (0.001) & 0.101 & 0.101 & 0.102 & 94.5 & -0.321 (0.002) & 0.153 & 0.152 & 0.151 & 95.1 & 0 (0.002) & 0.142 & 0.141 & 0.145 & 94.3 \\ 
1 & SPB & -0.16 & -0.126 (0.002) & 0.114 & 0.113 & 0.110 & 96.0 & 0 (0.001) & 0.105 & 0.104 & 0.105 & 95.4 & -0.164 (0.002) & 0.148 & 0.147 & 0.143 & 95.9 & 0 (0.002) & 0.136 & 0.136 & 0.137 & 95.2 \\ 
1 & SPB & 0 & 0.001 (0.002) & 0.119 & 0.119 & 0.114 & 95.7 & 0 (0.002) & 0.109 & 0.109 & 0.109 & 94.8 & 0.001 (0.002) & 0.143 & 0.143 & 0.137 & 95.7 & 0 (0.002) & 0.131 & 0.131 & 0.131 & 94.8 \\ [1ex]
1 & Simple & -0.32 & -0.226 (0.002) & 0.109 & 0.109 & 0.110 & 94.7 & -0.001 (0.001) & 0.101 & 0.101 & 0.103 & 94.7 & -0.318 (0.002) & 0.153 & 0.153 & 0.154 & 94.5 & -0.001 (0.002) & 0.142 & 0.142 & 0.145 & 94.6 \\ 
1 & Simple & -0.16 & -0.122 (0.002) & 0.114 & 0.113 & 0.113 & 95.2 & -0.001 (0.001) & 0.105 & 0.105 & 0.106 & 94.7 & -0.159 (0.002) & 0.148 & 0.148 & 0.148 & 95.1 & -0.001 (0.002) & 0.136 & 0.136 & 0.138 & 94.7 \\ 
1 & Simple & 0 & -0.002 (0.002) & 0.119 & 0.119 & 0.120 & 95.1 & -0.001 (0.002) & 0.109 & 0.109 & 0.110 & 94.9 & -0.003 (0.002) & 0.143 & 0.143 & 0.144 & 95.0 & -0.001 (0.002) & 0.131 & 0.131 & 0.132 & 94.8 \\ [2ex]
2 & SPB & -0.32 & -0.195 (0.002) & 0.111 & 0.110 & 0.108 & 95.3 & 0 (0.001) & 0.102 & 0.102 & 0.104 & 94.8 & -0.268 (0.002) & 0.151 & 0.151 & 0.149 & 95.3 & 0 (0.002) & 0.140 & 0.140 & 0.142 & 94.6 \\ 
2 & SPB & -0.16 & -0.108 (0.002) & 0.114 & 0.114 & 0.111 & 96.0 & 0 (0.001) & 0.105 & 0.105 & 0.106 & 95.5 & -0.138 (0.002) & 0.147 & 0.146 & 0.143 & 95.9 & 0 (0.002) & 0.135 & 0.135 & 0.136 & 95.3 \\ 
2 & SPB & 0 & 0.001 (0.002) & 0.119 & 0.119 & 0.114 & 95.7 & 0 (0.002) & 0.109 & 0.109 & 0.109 & 94.8 & 0.001 (0.002) & 0.143 & 0.143 & 0.137 & 95.7 & 0 (0.002) & 0.131 & 0.131 & 0.131 & 94.8 \\ [1ex]
2 & Simple & -0.32 & -0.193 (0.002) & 0.111 & 0.110 & 0.111 & 94.6 & -0.001 (0.001) & 0.102 & 0.102 & 0.105 & 94.8 & -0.265 (0.002) & 0.151 & 0.151 & 0.152 & 94.7 & -0.001 (0.002) & 0.140 & 0.140 & 0.143 & 94.8 \\ 
2 & Simple & -0.16 & -0.103 (0.002) & 0.115 & 0.114 & 0.114 & 95.3 & -0.001 (0.002) & 0.105 & 0.105 & 0.106 & 94.9 & -0.132 (0.002) & 0.147 & 0.147 & 0.147 & 95.1 & -0.001 (0.002) & 0.136 & 0.135 & 0.137 & 94.9 \\ 
2 & Simple & 0 & -0.002 (0.002) & 0.119 & 0.119 & 0.120 & 95.1 & -0.001 (0.002) & 0.109 & 0.109 & 0.110 & 94.9 & -0.003 (0.002) & 0.143 & 0.143 & 0.144 & 95.0 & -0.001 (0.002) & 0.131 & 0.131 & 0.132 & 94.8 \\ [2ex]
3 & SPB & -0.32 & -0.166 (0.002) & 0.111 & 0.111 & 0.108 & 95.7 & 0 (0.001) & 0.103 & 0.103 & 0.103 & 95.0 & -0.222 (0.002) & 0.149 & 0.149 & 0.144 & 95.5 & 0 (0.002) & 0.138 & 0.138 & 0.138 & 94.9 \\ 
3 & SPB & -0.16 & -0.088 (0.002) & 0.115 & 0.115 & 0.110 & 95.8 & 0 (0.001) & 0.106 & 0.106 & 0.105 & 95.1 & -0.112 (0.002) & 0.146 & 0.145 & 0.140 & 95.9 & 0 (0.002) & 0.134 & 0.134 & 0.134 & 95.1 \\ 
3 & SPB & 0 & 0.001 (0.002) & 0.119 & 0.119 & 0.114 & 95.7 & 0 (0.002) & 0.109 & 0.109 & 0.109 & 94.8 & 0.001 (0.002) & 0.143 & 0.143 & 0.137 & 95.7 & 0 (0.002) & 0.131 & 0.131 & 0.131 & 94.8 \\ [1ex]
3 & Simple & -0.32 & -0.170 (0.002) & 0.112 & 0.111 & 0.112 & 94.8 & -0.001 (0.001) & 0.103 & 0.103 & 0.104 & 95.0 & -0.227 (0.002) & 0.149 & 0.149 & 0.149 & 94.8 & -0.001 (0.002) & 0.138 & 0.138 & 0.139 & 95.1 \\ 
3 & Simple & -0.16 & -0.091 (0.002) & 0.115 & 0.115 & 0.116 & 95.0 & -0.001 (0.002) & 0.106 & 0.106 & 0.107 & 94.8 & -0.116 (0.002) & 0.146 & 0.146 & 0.147 & 94.9 & -0.001 (0.002) & 0.135 & 0.134 & 0.135 & 95.0 \\ 
3 & Simple & 0 & -0.002 (0.002) & 0.119 & 0.119 & 0.120 & 95.1 & -0.001 (0.002) & 0.109 & 0.109 & 0.110 & 94.9 & -0.003 (0.002) & 0.143 & 0.143 & 0.144 & 95.0 & -0.001 (0.002) & 0.131 & 0.131 & 0.132 & 94.8 \\ [2ex]
4 & SPB & -0.32 & -0.124 (0.002) & 0.113 & 0.113 & 0.109 & 95.8 & 0 (0.001) & 0.104 & 0.104 & 0.104 & 94.8 & -0.161 (0.002) & 0.147 & 0.147 & 0.142 & 95.8 & 0 (0.002) & 0.136 & 0.136 & 0.136 & 94.9 \\ 
4 & SPB & -0.16 & -0.064 (0.002) & 0.116 & 0.116 & 0.111 & 95.8 & 0 (0.002) & 0.107 & 0.106 & 0.106 & 95.1 & -0.081 (0.002) & 0.145 & 0.145 & 0.139 & 95.7 & 0 (0.002) & 0.133 & 0.133 & 0.134 & 95.0 \\ 
4 & SPB & 0 & 0.001 (0.002) & 0.119 & 0.119 & 0.114 & 95.7 & 0 (0.002) & 0.109 & 0.109 & 0.109 & 94.8 & 0.001 (0.002) & 0.143 & 0.143 & 0.137 & 95.7 & 0 (0.002) & 0.131 & 0.131 & 0.131 & 94.8 \\ [1ex]
4 & Simple & -0.32 & -0.127 (0.002) & 0.114 & 0.113 & 0.113 & 94.9 & -0.001 (0.001) & 0.105 & 0.104 & 0.105 & 95.1 & -0.165 (0.002) & 0.148 & 0.147 & 0.148 & 94.8 & -0.001 (0.002) & 0.136 & 0.136 & 0.136 & 95.0 \\ 
4 & Simple & -0.16 & -0.068 (0.002) & 0.116 & 0.116 & 0.117 & 94.9 & -0.001 (0.002) & 0.107 & 0.106 & 0.107 & 95.0 & -0.085 (0.002) & 0.145 & 0.145 & 0.146 & 95.0 & -0.001 (0.002) & 0.134 & 0.133 & 0.134 & 95.0 \\ 
4 & Simple & 0 & -0.002 (0.002) & 0.119 & 0.119 & 0.120 & 95.1 & -0.001 (0.002) & 0.109 & 0.109 & 0.110 & 94.9 & -0.003 (0.002) & 0.143 & 0.143 & 0.144 & 95.0 & -0.001 (0.002) & 0.131 & 0.131 & 0.132 & 94.8 \\ [2ex]
5 & SPB & -0.32 & -0.245 (0.001) & 0.081 & 0.081 & 0.075 & 96.3 & 0 (0.001) & 0.067 & 0.067 & 0.067 & 94.7 & -0.300 (0.001) & 0.099 & 0.099 & 0.093 & 96.4 & 0 (0.001) & 0.082 & 0.082 & 0.083 & 94.8 \\ 
5 & SPB & -0.16 & -0.123 (0.001) & 0.083 & 0.083 & 0.078 & 96.4 & 0 (0.001) & 0.070 & 0.070 & 0.070 & 94.8 & -0.139 (0.001) & 0.094 & 0.094 & 0.089 & 96.5 & 0 (0.001) & 0.079 & 0.079 & 0.080 & 95.0 \\ 
5 & SPB & 0 & 0.001 (0.001) & 0.086 & 0.086 & 0.080 & 96.4 & 0 (0.001) & 0.072 & 0.072 & 0.073 & 95.0 & 0.001 (0.001) & 0.091 & 0.091 & 0.085 & 96.5 & 0 (0.001) & 0.077 & 0.077 & 0.077 & 95.2 \\ [1ex]
5 & Simple & -0.32 & -0.244 (0.001) & 0.081 & 0.081 & 0.081 & 94.9 & 0.001 (0.001) & 0.067 & 0.067 & 0.068 & 94.4 & -0.298 (0.001) & 0.099 & 0.099 & 0.100 & 94.8 & 0.001 (0.001) & 0.082 & 0.082 & 0.084 & 94.9 \\ 
5 & Simple & -0.16 & -0.122 (0.001) & 0.083 & 0.083 & 0.083 & 95.1 & 0.001 (0.001) & 0.070 & 0.070 & 0.069 & 94.8 & -0.139 (0.001) & 0.095 & 0.094 & 0.094 & 95.2 & 0.001 (0.001) & 0.079 & 0.079 & 0.079 & 95.0 \\ 
5 & Simple & 0 & -0.001 (0.001) & 0.086 & 0.086 & 0.085 & 95.1 & 0.001 (0.001) & 0.072 & 0.072 & 0.072 & 94.8 & -0.001 (0.001) & 0.091 & 0.091 & 0.090 & 95.2 & 0.001 (0.001) & 0.077 & 0.077 & 0.077 & 94.9 \\ 
\bottomrule
\end{tabular}
\label{tab:auc_results_small}
\end{table}
\end{landscape}

\subsection{Distribution of Estimated Standard Errors\label{sec:simse}}

\begin{figure}[htbp]
    \centering
    \includegraphics[width=\textwidth]{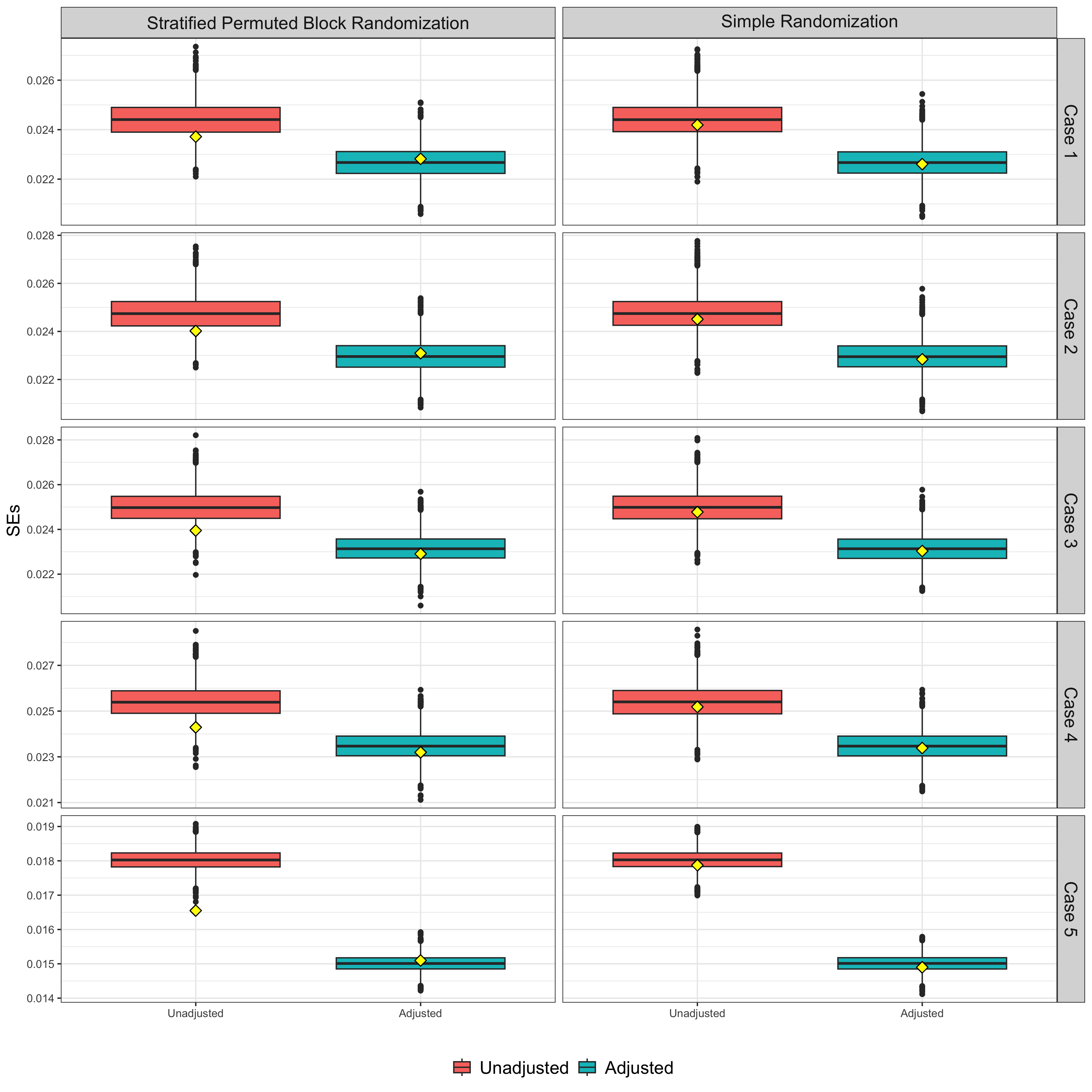}
    \caption{Estimated standard errors of $\sqrt{\hat{\sigma}_{L, \Delta}^{2}/n}$ (unadjusted) and $\sqrt{\hat{\sigma}_{CL, \Delta}^{2}/n}$ (adjusted) from 5,000 simulations under different randomization schemes and cases. Yellow diamonds represent the Monte Carlo standard deviations of 5,000 $\hat{\mathcal{U}}_{L,\Delta}$ (unadjusted) and $\hat{\mathcal{U}}_{CL,\Delta}$ (adjusted). The scalar $\theta$ is set to -0.32. $\tau$ is set to 2.}
    \label{se_dif}
\end{figure}

Figure \ref{se_dif} shows the distribution of estimated standard errors $\sqrt{\hat{\sigma}_{L, \Delta}^{2}/n}$ (unadjusted) and $\sqrt{\hat{\sigma}_{CL, \Delta}^{2}/n}$ (adjusted) from 5,000 simulations under different randomization schemes and cases, with $\theta=-0.32$ and $\tau=2$. Yellow diamonds represent the Monte Carlo standard deviations of 5,000 $\hat{\mathcal{U}}_{L,\Delta}$ (unadjusted) and $\hat{\mathcal{U}}_{CL,\Delta}$ (adjusted). 

For the unadjusted analysis under simple randomization and for both simple and SPB randomization in the adjusted analysis, the boxplots are narrow and include the yellow diamonds, despite some outliers. This observation confirms the accuracy of our variance formulas and validates the applicability of the covariate adjustment approach for both simple and covariate-adaptive randomization. In contrast, for the unadjusted analysis under SPB randomization, the boxplots are positioned above the corresponding yellow diamonds, indicating that the estimated standard errors overestimate the true standard deviations, resulting in conservative analyses.

\subsection{AUC vs LWYY}\label{sec:aucvslwyy}

\begin{figure}[htbp]
    \centering
    \includegraphics[width=\textwidth]{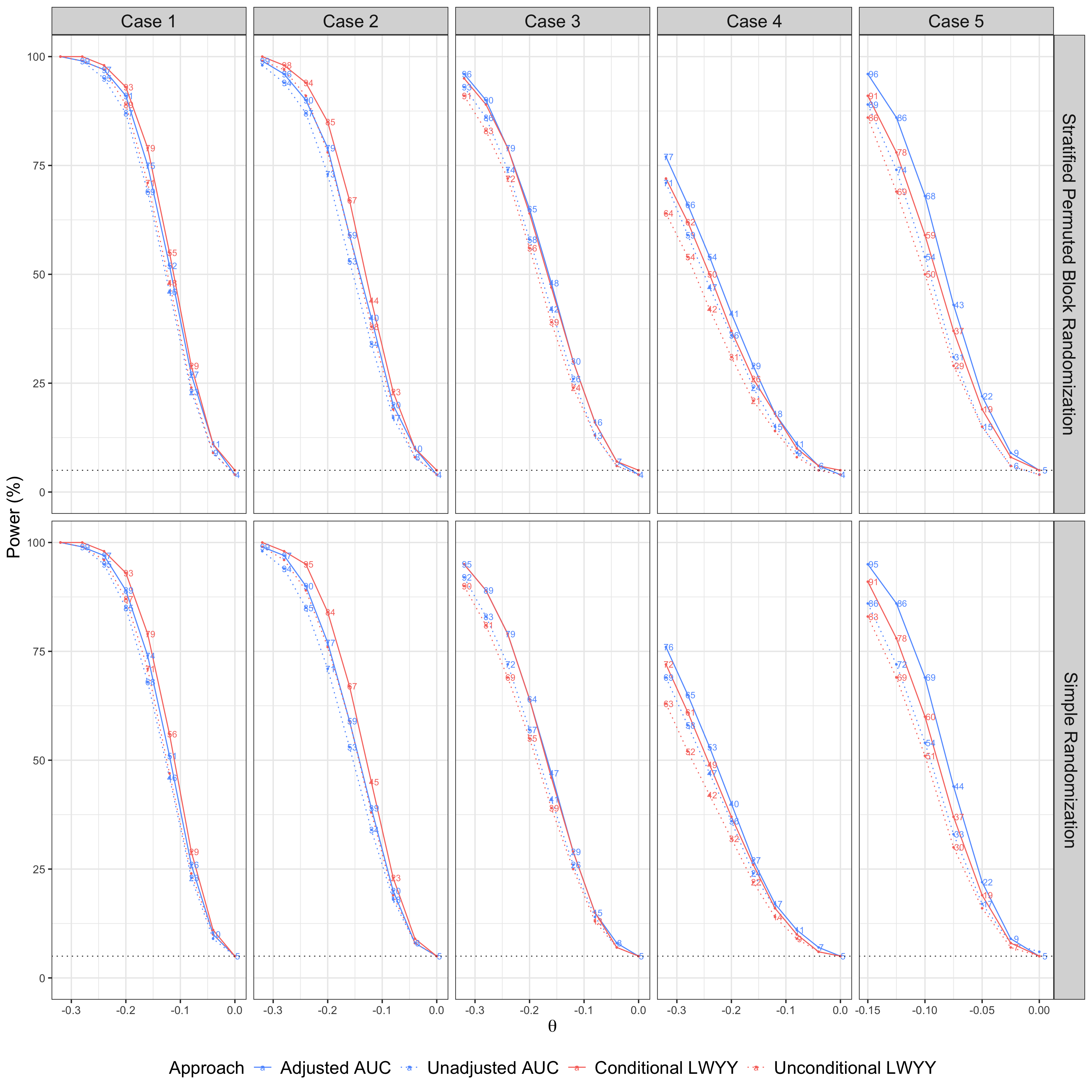}
    \caption{Comparison of power based on 5,000 simulations under different randomization schemes and cases. The dotted horizontal line indicates the significance level, $\alpha = 5\%$. $\tau$ is set to 2.}
    \label{power}
\end{figure}

One may be interested in the relative power between LWYY and AUC, although this is not the primary focus of this paper. Figure \ref{power} compares the power of AUC (both unadjusted and adjusted) with that of LWYY (both unconditional and conditional), acknowledging that the null hypotheses for AUC and LWYY differ, and that adjusted LWYY provides conditional estimates. The AUC is calculated over the interval $(0, \tau)$, whereas the LWYY is not limited to this interval and evaluates differences across the entire follow-up period. The unconditional LWYY estimate is derived from the model $\mu_{1}(t)=\mu_{0}(t)  \hbox{exp}(\theta)$, where $\mu_{1}(t)$ and $\mu_{0}(t)$ are the MCFs of the treatment and placebo arm, respectively. The conditional LWYY estimate is derived from the model $\mu(t \mid \boldsymbol{X}^{*})=\mu_{0}(t)  \hbox{exp}(\theta j + \boldsymbol{X}^{*T} \eta)$ for $j=0,1$. 

The relative power between AUC and LWYY is similar across different randomization schemes but varies across cases. We focus on the comparison between conditional LWYY and adjusted AUC. In Case 1, where the LWYY model is correctly specified, the LWYY model has higher power because its inference is based on likelihood, providing optimal power. In Case 2, where the treatment effect is larger at later times, although the LWYY model is misspecified, it still has better power than AUC. In Case 3, the adjusted AUC has comparable power to the conditional LWYY, but the unadjusted AUC has slightly better power than the unconditional LWYY. In Cases 4 and 5, where the treatment effect is larger at earlier times, AUC has better power than LWYY.

These results suggest that AUC may provide better power than LWYY if the treatment effect decreases over time. This is because the test statistic of AUC assigns lesser weights to later differences compared to the LWYY model \citep{tian2018efficiency, uno2023ratio}. Therefore, when the treatment effect is larger at later study time points, AUC tends to be less powerful than LWYY. Conversely, if the treatment effect is smaller at later study time points, AUC is likely to be more powerful than LWYY. In the FINEARTS-HF study, the treatment effects on mean cumulative events appear to diminish over time, resembling Case 4 in our simulations, where AUC might demonstrate greater power than LWYY.

From an interpretative perspective, in Cases 2 to 5, where the treatment effect varies over time, the AUC provides a more clinically meaningful interpretation, as the LWYY model assumes and estimates a constant treatment effect. Furthermore, the adjusted AUC preserves the unconditional interpretation, while the adjusted LWYY targets a conditional estimand.

The choice of $\tau$ may change the relative power between LWYY and AUC. With a smaller $\tau$, AUC counts a smaller number of recurrent events, which might suggest diminished power for AUC. One might then question whether AUC can still have better power than LWYY with a smaller $\tau$. Interestingly, Table \ref{powertable} shows that in Cases 4 and 5, AUC even has better power with $\tau=1.7$ than with $\tau=2$. This is because, in these cases, setting a smaller $\tau$ ignores the time period when the treatment effect is relatively smaller, so $\hat{U}_1(1.7) / \hat{U}_0(1.7) < \hat{U}_1(2) / \hat{U}_0(2)$, as shown in Figure 1. These results suggest that in cases where the treatment effect decreases over time, AUC with a smaller $\tau$ is likely to yield even higher power.

\begin{table}[htbp]
\centering
\caption{Power of AUC in Cases 4 and 5 with different choices of $\tau$ under various randomization schemes.}
\label{powertable}
\begin{tabular}{ccccccc}
\toprule
 &  &  & \multicolumn{4}{c}{Power (\%)} \\ 
\cmidrule(lr){4-7}
\multicolumn{3}{c}{Settings} & \multicolumn{2}{c}{$\tau = 2$} & \multicolumn{2}{c}{$\tau = 1.7$} \\ 
\cmidrule(lr){1-3} \cmidrule(lr){4-5} \cmidrule(lr){6-7} 
Cases & Schemes & $\theta$ & Unadjusted & Adjusted & Unadjusted & Adjusted \\ \midrule 
4 & SPB & -0.32 & 71 & 77 & 74 & 79 \\ 
4 & SPB & -0.28 & 59 & 66 & 63 & 68 \\ 
4 & SPB & -0.24 & 47 & 54 & 50 & 56 \\ 
4 & SPB & -0.20 & 36 & 41 & 38 & 43 \\ 
4 & SPB & -0.16 & 24 & 29 & 26 & 30 \\ 
4 & SPB & -0.12 & 15 & 18 & 16 & 19 \\ 
4 & SPB & -0.08 & 9 & 11 & 10 & 11 \\ 
4 & SPB & -0.04 & 5 & 6 & 5 & 6 \\ 
4 & SPB & 0 & 4 & 4 & 4 & 5 \\ 
4 & Simple & -0.32 & 69 & 76 & 72 & 78 \\ 
4 & Simple & -0.28 & 58 & 65 & 61 & 68 \\ 
4 & Simple & -0.24 & 47 & 53 & 50 & 55 \\ 
4 & Simple & -0.20 & 36 & 40 & 38 & 42 \\ 
4 & Simple & -0.16 & 24 & 27 & 26 & 29 \\ 
4 & Simple & -0.12 & 16 & 17 & 17 & 18 \\ 
4 & Simple & -0.08 & 10 & 11 & 10 & 11 \\ 
4 & Simple & -0.04 & 6 & 7 & 6 & 7 \\ 
4 & Simple & 0 & 5 & 5 & 5 & 5 \\ 
5 & SPB & -0.15 & 89 & 96 & 91 & 97 \\ 
5 & SPB & -0.125 & 74 & 86 & 77 & 89 \\ 
5 & SPB & -0.10 & 54 & 68 & 56 & 71 \\ 
5 & SPB & -0.075 & 31 & 43 & 33 & 46 \\ 
5 & SPB & -0.05 & 15 & 22 & 16 & 23 \\ 
5 & SPB & -0.025 & 6 & 9 & 6 & 9 \\ 
5 & SPB & 0 & 4 & 5 & 3 & 5 \\ 
5 & Simple & -0.15 & 86 & 95 & 88 & 96 \\ 
5 & Simple & -0.125 & 72 & 86 & 75 & 88 \\ 
5 & Simple & -0.10 & 54 & 69 & 56 & 72 \\ 
5 & Simple & -0.075 & 33 & 44 & 35 & 47 \\ 
5 & Simple & -0.05 & 17 & 22 & 18 & 23 \\ 
5 & Simple & -0.025 & 8 & 9 & 8 & 9 \\ 
5 & Simple & 0 & 6 & 5 & 5 & 5 \\ 
\bottomrule
\end{tabular}
\end{table}

\section{Covariate adjustment for RMST}\label{sec:rmst}

In this section, we reuse many notations from the main text, originally defined in the context of the area under the mean cumulative function, but redefine them in the context of the area under the survival function.

\subsection{Data and Estimand}

Let $D$ represent the time to a terminal event and $C$ denote the independent censoring time. An observation consists of $\{T, \delta \}$, where $T=C \wedge D$, $\delta=\mathbb{I}(D \leq C)$, and $a \wedge b =\hbox{min}(a, b)$. The observed data set $\mathcal{D}=\{T_{i}, \delta_{i} \}$, for $i=1, \dots, n$, consists of independent realizations of $\{T, \delta \}$. Let $S(u)=\mathbb{P}(D \geq u)$ denote the survival function. Reusing notations from the main text, we define the area under the survival curve $S(t)$ up to time $\tau$ as $U(\tau)=\int_{0}^{\tau}S(u)du$. Additionally, let $\hat U(\tau)=\int_{0}^{\tau} \hat S(u)du$.

Define the terminal event martingale as $M_i^{D}(t)=N_{i}^{D}(t) - \int_{0}^{t} \mathbb{I}_{i}(T \geq u) dA^{D}(u) $, where $N^D(t) = I(D \leq t)$ is the counting process for $D$ and $dA^{D}(u) = \mathbb{E}\{ dN^{D}(u) \mid T \geq u \}$. Then, we have:
\begin{eqnarray*}
    \hat U(\tau) - U(\tau) &=& \int_{0}^{\tau} \left\{ \hat S(u) - S(u)\right\} du \approx \int_{0}^{\tau} -\hat S(u)  \left\{ \hat \Lambda(u) - \Lambda(u) \right\} du \approx \int_{0}^{\tau} -\hat S(u)  \frac{1}{n} \sum_{i=1}^{n} \int_{0}^{u} \frac{d\hat M_i^D(s)}{\hat S(s)} du \\
    &=& -\frac{1}{n} \sum_{i=1}^{n} \int_{0}^{\tau} \int_{0}^{u} I(s \leq u)  \hat S(u)  \frac{d\hat M_i^D(s)}{\hat S(s)} du = -\frac{1}{n} \sum_{i=1}^{n}  \int_{0}^{\tau} \left\{\int_{s}^{\tau} \hat S(u) \, du \right\}  \frac{d\hat M_i^D(s)}{\hat S(s)}.
\end{eqnarray*}
where $\Lambda(u)$ is the cumulative hazard function, $\hat \Lambda(u)$ is its estimate, $d \hat A^{D}(u) = \sum_{i=1}^{n} dN^{D}_{i}(u)/\sum_{i=1}^{n} \mathbb{I}(T_i \geq u)$, and $d\hat M_{i}^{D}(u) = dN_{i}^{D}(u) - \mathbb{I}_{i}(T \geq u) d \hat A^{D}(u)$.

The first approximation arises from the Taylor expansion of $\hat S(u)=e^{-\hat \Lambda(u)}$, while the second is based on the martingale expansion formula for the Nelson-Aalen estimator.

Reusing notations from the main text, we define $\hat \psi_{i}(\tau)=-\int_{0}^{\tau} \left\{\int_{s}^{\tau} \hat S(u) du \right\}  \frac{d\hat M_i(s)}{\hat S(s)}$. Thus, we have $\sqrt{n} \{ \hat U(\tau) - U(\tau)  \} = \frac{1}{\sqrt{n}} \sum_{i=1}^{n} \hat \psi_{i}(\tau) +o_p(1)$, where $o_p(1)$ denotes a term tending to 0 in probability as the sample size $n \to \infty$.

Denote $U_0(\tau)$ and $U_1(\tau)$ as the RMST for placebo and treatment arms. The treatment effect can be quantified as either the difference, $\Delta U(\tau)=U_1(\tau) - U_0(\tau)$, or the ratio, $U_1(\tau)/U_0(\tau)$. 

\subsection{Nonparametric Covariate Adjustment}
The core concept involves linearizing the test statistic to obtain an outcome for each patient, followed by the application of generalized regression adjustment or augmentation to these outcomes \citep{zhang2008improving, tsiatis2008covariate}. In this discussion, we focus on the difference in RMST. Covariate adjustment for the ratio of RMST can be derived similarly.

Let $\Delta \hat U(\tau) = \hat U_1(\tau) - \hat U_0(\tau)$ and $ \hat{\mathcal{U}}_{L, \Delta}=\frac{\sqrt{n_0}\sqrt{n_1}}{n} \left\{ \Delta \hat U(\tau) - \Delta U(\tau) \right\} $, where $n_0$ and $n_1$ represent the sample sizes for the placebo and treatment arms, respectively. The linearization of $\hat{\mathcal{U}}_{L,\Delta}$ is as follows:
\begin{eqnarray*}
    \hat{\mathcal{U}}_{L,\Delta} &=& \frac{\sqrt{n_0}\sqrt{n_1}}{n} \left\{ \hat U_{1}(\tau) -  U_{1}(\tau) \right\} - \frac{\sqrt{n_0}\sqrt{n_1}}{n} \left\{  \hat U_{0}(\tau) - U_{0}(\tau) \right\} \\
    &=& \frac{1}{n} \left( \frac{n_0}{n_1} \right)^{\frac{1}{2}} \sum_{i=1}^{n_1} \hat \psi_{i}(\tau) - \frac{1}{n} \left( \frac{n_1}{n_0} \right)^{\frac{1}{2}} \sum_{j=1}^{n_0} \hat \psi_{j}(\tau) +o_p(1) \\
    &=& \frac{1}{n} \sum_{i=1}^{n} \left\{ I_{i}  \left( \frac{n_0}{n_1} \right)^{\frac{1}{2}}   \hat \psi_{i}(\tau) - (1-I_i)  \left( \frac{n_1}{n_0} \right)^{\frac{1}{2}}   \hat \psi_{i}(\tau) \right\} +o_p(1) \\
    &=& \frac{1}{n} \sum_{i=1}^{n} \left\{ I_{i}  \hat \psi_{i}^{\Delta}(\tau) - (1-I_i)  \hat \psi_{i}^{\Delta}(\tau) \right\} +o_p(1)
\end{eqnarray*}
where $I_{i}=1$ if subject $i$ is in the treatment arm and 0 otherwise, and $\hat \psi_{i}^{\Delta}(\tau) = \left( \frac{n_1}{n_0} \right)^{\frac{1}{2}  (-1)^{I_i}} \hat \psi_{i}(\tau)$. The unadjusted test statistic for the difference in AUCs is given by $\mathcal{T}_{L,\Delta} = \sqrt{n}  \hat{\mathcal{U}}_{L, \Delta}/\hat{\sigma}_{L, \Delta}$, where 
$$\hat{\sigma}_{L, \Delta}^{2} = \frac{1}{n}  \left[ n_0  \mathbb{E}\left\{ \hat \psi_{1}(\tau)^{T} \hat \psi_{1}(\tau) \right\} + n_1  \mathbb{E}\left\{ \hat \psi_{0}(\tau)^{T} \hat \psi_{0}(\tau)\right\} \right] $$
Here, $\hat \psi_{1}(\tau)$ and $\hat \psi_{0}(\tau)$ consist of $\hat \psi_{i}(\tau)$ from subjects in the treatment and placebo arms, respectively. 

Under simple randomization, under the null hypothesis $H_{0, \Delta}: \Delta U(\tau)=0$ or the alternative hypothesis, we have $\sqrt{n}  \hat{\mathcal{U}}_{L,\Delta}\stackrel{d}{\to} N\left(0, \sigma_{L, \Delta}^{2}\right) $, $\hat{\sigma}_{L, \Delta}^{2} \stackrel{p}{\to} \sigma_{L, \Delta}^{2}$, and $\mathcal{T}_{L, \Delta} \stackrel{d}{\to} N(0, 1) $.

Let $\boldsymbol{X}^{*}$ be a $p$-dimensional vector of covariates. Define $X_{i}=X_{i}^{*}-\bar X$, so that $\mathbb{E}(X_i)=0$. Following \cite{ye2024covariate}, we treat $\hat \psi_{i}^{\Delta}(\tau)$ as the outcome and apply the generalized regression adjustment or augmentation \citep{tsiatis2008covariate}. This results in the covariate-adjusted $\hat{\mathcal{U}}_{CL,\Delta}$:
\begin{eqnarray*}
    \hat{\mathcal{U}}_{CL,\Delta} &=& \frac{1}{n} \sum_{i=1}^{n} \left\{ I_{i}   (\hat \psi_{i}^{\Delta}(\tau) - X_{i}^{T} \hat \beta_{1}^{\Delta} )  - (1-I_i)  (\hat \psi_{i}^{\Delta}(\tau) - X_{i}^{T} \hat \beta_{0}^{\Delta})  \right\}  = \hat{\mathcal{U}}_{L,\Delta} - \hat{\mathcal{A}}_{\Delta}
\end{eqnarray*}
where $\hat{\mathcal{A}}_{\Delta} = \frac{1}{n} \sum_{i=1}^{n} \left\{ I_i  X_{i}^{T}  \hat \beta_{1}^{\Delta}  - (1-I_i)  X_{i}^{T}  \hat \beta_{0}^{\Delta} \right\}$, $\hat \beta_{j}^{\Delta} = \left( \sum_{i:I_i=j} X_i X_{i}^{T} \right)^{-1} \sum_{i:I_i=j} X_{i} \hat \psi_{i}^{\Delta}(\tau) $.

Let $\Delta \hat U^{(a)}(\tau)$ denote the adjusted $\Delta \hat U(\tau)$, and is given by $\Delta \hat U^{(a)}(\tau)=\Delta \hat U(\tau) - \frac{n}{\sqrt{n_0}\sqrt{n_1}} \hat{\mathcal{A}}_{\Delta}$. Then $ \hat{\mathcal{U}}_{CL, \Delta}=\frac{\sqrt{n_0}\sqrt{n_1}}{n} \{ \Delta \hat U^{(a)}(\tau) - \Delta U(\tau) \} $.

Define the covariate-adjusted test statistic for the difference in AUCs as $\mathcal{T}_{CL,\Delta} = \sqrt{n}  \hat{\mathcal{U}}_{CL, \Delta}/\hat{\sigma}_{CL, \Delta}$, where $\hat{\sigma}_{CL, \Delta}^2= \hat{\sigma}_{L, \Delta}^2 - \frac{n_0 n_1}{n^2} (\hat \beta_{1}^{\Delta} + \hat \beta_{0}^{\Delta})^{T}   \sum_{X}  (\hat \beta_{1}^{\Delta} + \hat \beta_{0}^{\Delta}) $. 

The following results hold under both simple randomization and covariate-adaptive randomization schemes: Under the $H_{0, \Delta}: \Delta U(\tau)=0$ or the alternative hypothesis, $\sqrt{n}  \hat{\mathcal{U}}_{CL,\Delta}\stackrel{d}{\to} N\left(0, \sigma_{CL, \Delta}^{2}\right) $, $\hat \sigma_{CL, \Delta}^{2} \stackrel{p}{\to} \sigma_{CL, \Delta}^{2}$, and $\mathcal{T}_{CL, \Delta} \stackrel{d}{\to} N(0, 1) $.

Thus, $H_{0,\Delta}$ is rejected if and only if $|\mathcal{T}_{CL,\Delta}| > z_{\alpha/2}$. The adjusted $100(1-\alpha) \%$ CI for $\Delta U(\tau)$ is $( \Delta \hat U^{(a)}(\tau) - z_{\alpha/2}  SE\left\{\Delta \hat U^{(a)}(\tau)\right\}, \Delta \hat U^{(a)}(\tau) + z_{\alpha/2}  SE\left\{\Delta \hat U^{(a)}(\tau)\right\} )$, where $SE\left\{ \Delta \hat U^{(a)}(\tau)\right\} = \left( \frac{n}{n_0 n_1}  \hat{\sigma}_{CL, \Delta}^2  \right)^{1/2}$.

\subsection{Simulations}
\subsubsection{Data Generation}
Let $\boldsymbol{X}=(X_1, X_2, X_3)$ be a 3-dimensional covariate vector following the 3-dimensional mutually independent standard normal distribution. We consider both stratified permuted block (SPB) randomization and simple randomization in this context. In SPB randomization, subjects are categorized into strata based on $X_1$ and $X_2$, where $X_2$ is categorized into four quantiles. Within each stratum, subjects are further grouped into blocks of size 4, and permuted block randomization is applied within these blocks.

For our simulation, we have a total of 5,000 subjects, using two scenarios from the work of \cite{ye2024covariate}:

\begin{itemize}
    \item Case 1: The conditional hazard follows a Cox model, $\lambda_j(t|\boldsymbol{X})=\{\hbox{log}(2)\}  \hbox{exp}(-\theta j + \eta^{T}\boldsymbol{X})$ for $j=0,1$, where $\theta$ denotes a scalar parameter, $\eta = (0.5, 0.5, 0.5)$. The censoring variables follow uniform distribution on interval $(10, 40)$ and are independent of $\boldsymbol{X}$.
    \item Case 2: $D_j=\hbox{exp}(\theta j +\eta^{T}\boldsymbol{X})+\epsilon$, $j=0,1$, where $\theta$, $\eta$ and $\boldsymbol{X}$ are the same as those in case 1, and $\epsilon$ is a random variable independent of censoring and $\boldsymbol{X}$ and has the standard exponential distribution. The setting for censoring is the same as that in case 1.
\end{itemize}

\subsubsection{Results}

\begin{figure}[p]
    \centering
    \includegraphics[width=1\textwidth]{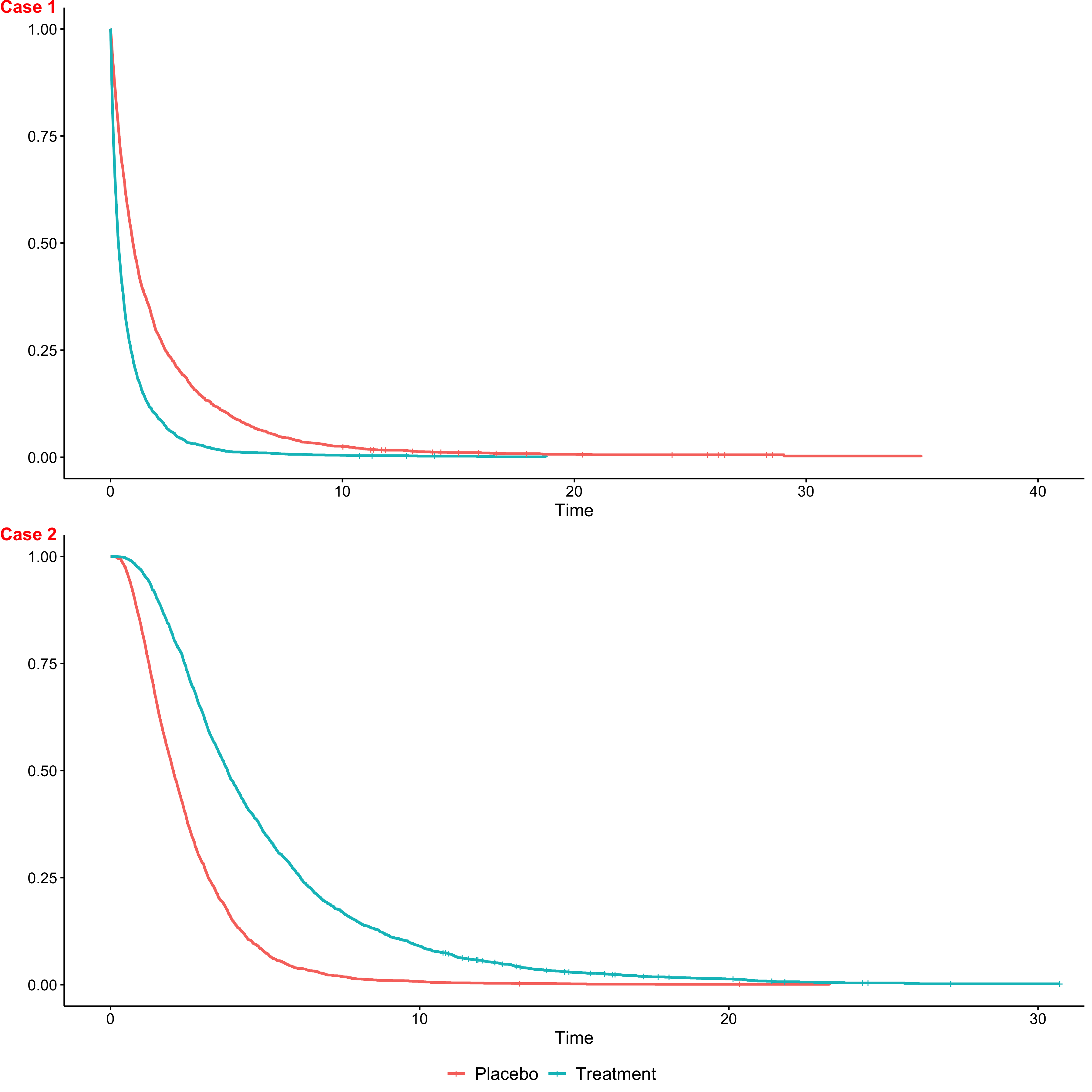}
    \caption{Survival curves for two simulation cases using the simple randomization scheme. Each case is based on data from a single simulation with a total sample size of 5,000 (allocated in a 1:1 ratio). The scalar $\theta$ is set to 1.}
    \label{kmcurve}
\end{figure}

Figure \ref{kmcurve} displays the survival curves for the two cases under the simple randomization scheme, utilizing a total sample size of 5,000 with equal allocation. The scalar parameter $\theta$ is set to 1.

\begin{figure}[p]
    \centering
    \includegraphics[width=1\textwidth]{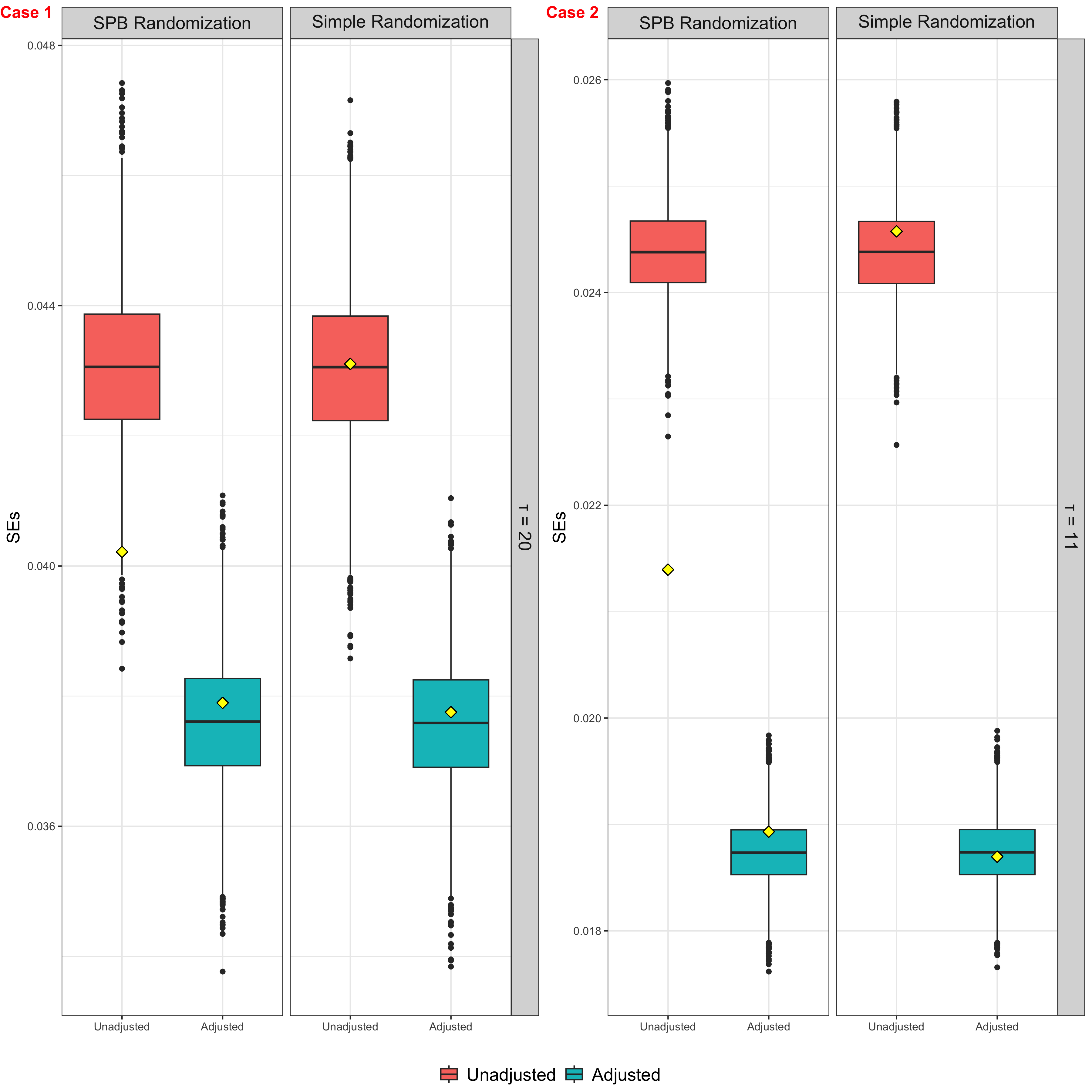}
    \caption{Estimated standard errors of $\sqrt{\hat{\sigma}_{L, R}^{2}/n}$ (unadjusted) and $\sqrt{\hat{\sigma}_{CL, R}^{2}/n}$ (adjusted) from 5,000 simulations under different randomization schemes and cases. Yellow diamonds represent the Monte Carlo standard deviations of 5,000 $\hat{\mathcal{U}}_{L,R}$ (unadjusted) and $\hat{\mathcal{U}}_{CL,R}$ (adjusted). The scalar $\theta$ is set to 0.}
    \label{se_rmst}
\end{figure}

Figure \ref{se_rmst} shows the distribution of estimated standard errors $\sqrt{\hat{\sigma}_{L, \Delta}^{2}/n}$ (unadjusted) and $\sqrt{\hat{\sigma}_{CL, \Delta}^{2}/n}$ (adjusted) from 5,000 simulations under different randomization schemes and cases, with $\theta=0$. The RMST is calculated over the interval $(0, \tau)$. The results for $\sqrt{\hat{\sigma}_{L, R}^{2}/n}$ and $\sqrt{\hat{\sigma}_{CL, R}^{2}/n}$ are similar. The yellow diamonds represent the Monte Carlo standard deviations of 5,000 $\hat{\mathcal{U}}_{L, \Delta}$ (unadjusted) and $\hat{\mathcal{U}}_{CL, \Delta}$ (adjusted). 

In the context of simple randomization for the unadjusted analysis, as well as for both simple and SPB randomization in the adjusted analysis, the boxplot bodies are thin and include the yellow diamond, despite some outliers. This empirically supports the accuracy of our variance formulas and the applicability of the covariate adjustment approach to both simple and covariate-adaptive randomization. In contrast, for SPB randomization, the boxplots from the unadjusted analysis are positioned above the corresponding yellow diamonds, indicating that the estimated standard errors tend to overestimate the actual standard deviations, leading to conservative analyses.

\begin{figure}[p]
    \centering
    \includegraphics[width=1\textwidth]{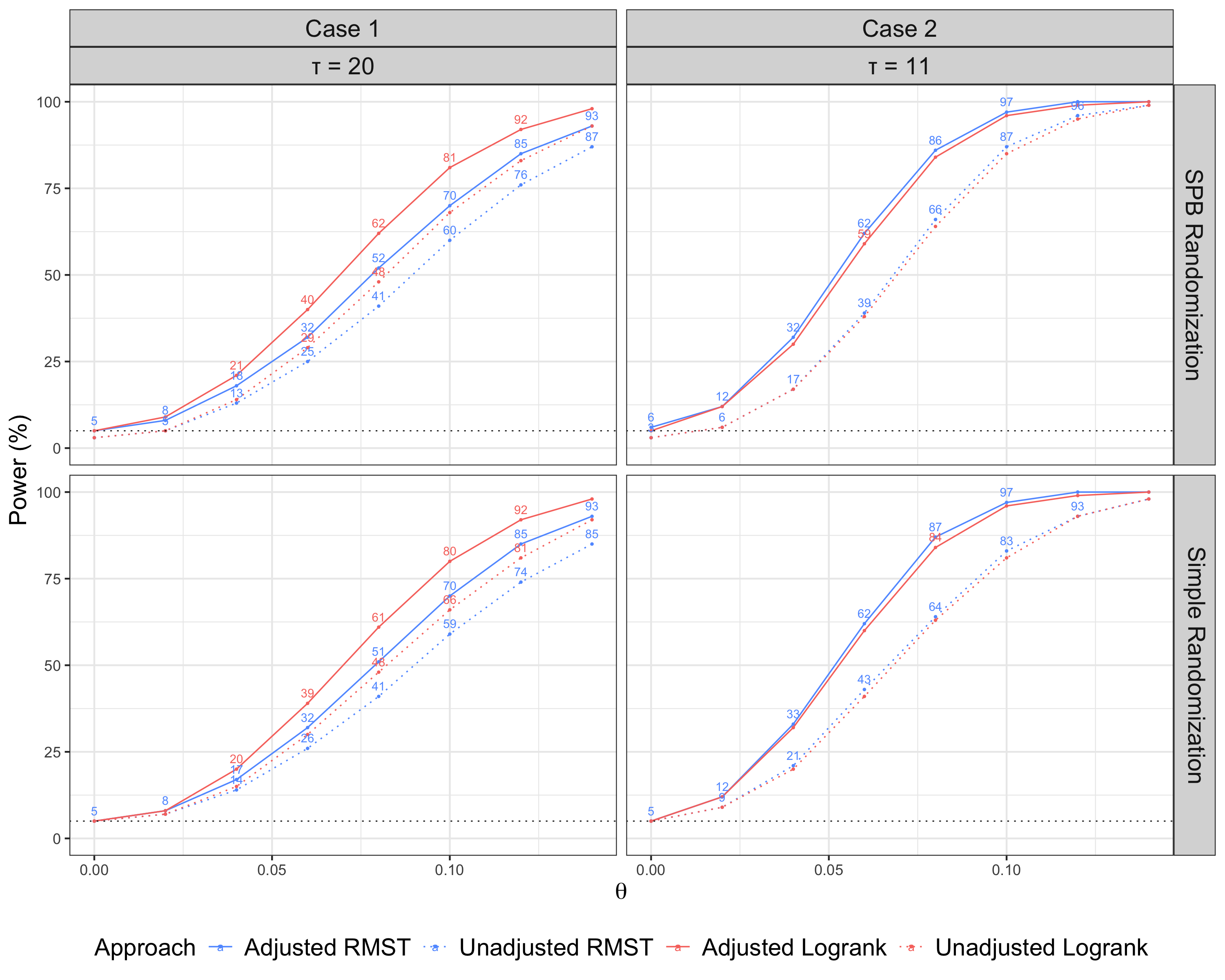}
    \caption{Comparison of power across various approaches based on 5,000 simulations under different randomization schemes and cases. The dotted horizontal line indicates the significance level, $\alpha = 5\%$.}
    \label{power_rmst}
\end{figure}

Figure \ref{power_rmst} compares the power of unadjusted RMST with adjusted RMST based on 5,000 simulations. Tests on the difference and ratio of RMSTs show identical power. The figure illustrates that the adjusted analysis consistently achieves higher power across various cases and randomization schemes. Additionally, under SPB randomization, the Type I errors for the unadjusted analysis fall below the nominal level, suggesting that failing to account for covariate-adaptive randomization results in conservative estimates.

Furthermore, one might be interested in the relative power between the log-rank test and RMST. Therefore, Figure \ref{power_rmst} also contrasts the power of RMST with that of the log-rank test, while acknowledging that the null hypotheses of RMST and log-rank test are different. The adjusted log-rank test follows the approach proposed by \cite{ye2024covariate}. The RMST is calculated over the interval $(0, \tau)$, whereas the log-rank test is not limited to this interval and evaluates survival differences across the entire follow-up period. In Case 1, where the proportionality assumption holds, the log-rank test exhibits higher power. Conversely, in Case 2, where the proportionality assumption is violated, RMST demonstrates slightly better power than the log-rank test.

\bibliography{supplementary}